\documentclass[aps,prl,twocolumn,floatfix,superscriptaddress]{revtex4}

\usepackage{amssymb,amsmath,amstext}                
\usepackage{algorithm}
\usepackage{algpseudocode}
\usepackage{graphicx}
\usepackage{epstopdf}                                               
\usepackage{color}                                                     
\usepackage{bm}                                                        
\usepackage{appendix}

\usepackage[utf8]{inputenc}
\usepackage{lipsum}

\usepackage{ulem}
\usepackage{latexsym}
\usepackage[colorlinks=true,citecolor=blue,linkcolor=magenta]{hyperref}
\usepackage{soul}
\usepackage{amsmath}
\usepackage{float}
\usepackage{placeins}

\def\be{\begin{equation}}
\def\ee{\end{equation}}
\def\bes{\begin{equation*}}
\def\ees{\end{equation*}}

\def\bea{\begin{eqnarray}}
\def\eea{\end{eqnarray}}

\def\beas{\begin{eqnarray*}}
\def\eeas{\end{eqnarray*}}

\def\bi{\begin{itemize}}
\def\ei{\end{itemize}}

\definecolor{dgreen} {RGB}{78,138,21}

\definecolor{giergiel} {RGB}{0,128,88}

\definecolor{purple} {RGB}{128,0,160}

\usepackage{orcidlink}

\begin{document}
\title{Hybrid Quantum-Classical Scheduling with Problem-Aware Calibration on a Quantum Annealer}
\author{Krzysztof Giergiel\orcidlink{0000-0003-3297-796X}}
\thanks{krzysztof.giergiel@csiro.au}
\affiliation{CSIRO, Manufacturing, Private Bag 10, Clayton, VIC 3169, Australia}
\author{Y. Sam Yang\,\orcidlink{0000-0002-8565-6214}}
\affiliation{CSIRO, Manufacturing, Private Bag 10, Clayton, VIC 3169, Australia}
\author{Anthony B. Murphy\,\orcidlink{0000-0002-2820-2304}}
\affiliation{CSIRO, Manufacturing, PO Box 218, Lindfield NSW 2070, Australia}

\begin{abstract}
We evaluate the application of quantum annealing (QA) to a real-world combinatorial optimisation problem---room scheduling for sports camps at the Australian Institute of Sport---using both classical and quantum approaches. Due to current hardware limitations, the full problem cannot be embedded on existing QA platforms, motivating a hybrid method combining classical heuristics with quantum subroutines. We develop an improved formulation and a novel problem-aware calibration scheme, leveraging multi-qubit statistics to enhance the annealing performance. Our results show that the optimal annealing time aligns with the coherence time of the D-Wave Advantage 2 prototype ($\approx 100~ns$), with longer anneals yielding poorer outcomes before slow thermal recovery. Despite calibration and formulation improvements, QA performance degrades with increased connectivity and problem size, highlighting the need for improved qubit quality and parameter precision. These findings clarify the capabilities and limitations of current QA hardware and suggest strategies for extending its practical utility through hybrid methods and informed calibration.
\end{abstract}
\date{\today}

\maketitle
\section{Introduction}

Optimization problems appear in a wide range of scientific, industrial, and engineering applications~\cite{Nocedal2006}. Approaches to solving them include exact methods (e.g., integer or convex programming), classical heuristics such as simulated annealing~\cite{Vecchi}, and more recent methods like quantum annealing~\cite{Kadowaki1998}. Typically, the goal is to find the global minimum of a cost function, but this is usually not guaranteed, especially for non-convex or combinatorial problems~\cite{Handbook}. The likelihood of success, computational efficiency, and scalability vary depending on both the algorithmic approach and problem structure, making the optimal choice of method highly problem-dependent.


Quantum annealing (QA) is a deterministic quantum algorithm designed to find the global minimum of a problem's cost function by evolving a quantum system toward the ground state of a problem Hamiltonian. It operates based on the following time-dependent Hamiltonian: 
\bea
H(s) = (1-s)X + sZ,\quad 
\\X = \sum X_i,\ Z = \sum Z_i + \sum_{\langle i,j \rangle} Z_i Z_j,
\eea
where the interpolation parameter \( s \in [0, 1] \) is increased during the annealing. The system is prepared in a state fully magnetised in the $X$ direction and provided $s$ is changing slowly enough, this state will be interpolated to the ground state of the optimisation problem encoded by the $Z$ magnetic fields and interaction terms, with $Z_i$ proportional to a Pauli $\sigma_z$ operator acting on spin $i$, with the proportionality coefficients encoding the optimisation problem. Each individual $X_i$ term is proportional to a Pauli $\sigma_x$ operator acting on spin $i$ and from the point of view of optimisation problem can be approximately understood as a bit flip. As $s$ increases, the influence of the bit-flip terms diminishes relative to the energy term, illustrating the connection to simulated annealing. 

The expected speed-up over direct enumeration from a perfectly functioning large quantum annealer is similar to the Grover algorithm: a square root of the number of enumerated elements \cite{Jorg2009,Pujos2018,Lidar2018}. However, it is debated whether QA can achieve even this limit\cite{Sondhi}. A fundamental limitation of QA is the adiabatic condition, related to the Landau-Zener non-adiabatic transition probability. This involves transitioning to an excited state instead of the ground state (global minimum) and implies that the annealing time must increase as the minimum energy gap decreases. Quantum annealing is anticipated to have exponential runtime for NP-hard problems, such as general optimisation, where the minimum energy gap decreases exponentially with the problem size\cite{Smelyanskiy,HenYoung,Jrg2009EnergyGI}. 

Comparing QA to direct enumeration is problematic because many problems have inherent structure. For example, in the NP-Hard subset sum problem with a set size $N$, direct enumeration requires exploring $2^N$ combinations, whereas the most efficient classical algorithm has a runtime of $O(2^{0.291N})$\cite{Joux}, which is significantly better then $t_a$ =$2^{(N/2)}$ expected for non-specific Grover speed-up.

Another challenge with QA is that readout and thermal fluctuations introduce noise into the computation results. Typically, only fluctuations occurring after crossing the minimal gap will produce local bit flip errors and ones before can produced states very different from the ground state. These small bit-flip distance errors have to be mitigated with classical post-processing to satisfactory results\cite{Pelofske2023}. 

A final problem with the QA platform is small range of model parameters. These parameters are steered by analogue voltages limiting the precision available. Even if perfect digital-analog conversion was possible the system would be limited by its finite temperature, which is 0.01 of current highest accessible $Z_i$ values.

Despite these drawbacks of quantum annealing, it is being applied to a wide range of optimisation problems with varying degrees of success. Several applications are routinely implemented on the D-Wave QAs. On the other hand, the physical limitations noted above limit the usefulness of QA, and  the wide range of classical optimisation approaches available make definitive comparisons difficult. Therefore, it is valuable to assess the application of QA to a real-world problems, understand its limitations, and benchmark its performance using quantum and classical approaches.

In this article we first introduce the problem of room scheduling optimisation for sports camps at the Australian Institute of Sport (AIS) using quantum and classical computing approaches. We explore exact and heuristic classical solutions, as well as quantum annealing methods using the D-Wave Advantage 2 prototype. We assesses the feasibility of quantum annealing for this problem, considering current hardware limitations and the embedding challenges posed by high connectivity requirements. We also propose an improved formulation that reduces parameter range constraints and enhances the embedding efficiency on quantum annealing hardware. We introduce a new calibration procedure, where we calibrate multi-qubit statistics relevant to the optimisation problem. This is in contrast to the normal single-qubit calibration. 
The results provide insights into future scalability and the potential role of quantum computing in combinatorial optimization problems, which exist in real world applications.

\section{Room scheduling optimization at AIS campus}
\subsection{The full problem}
\begin{figure*}[!tp]
\includegraphics[width=0.99\textwidth]{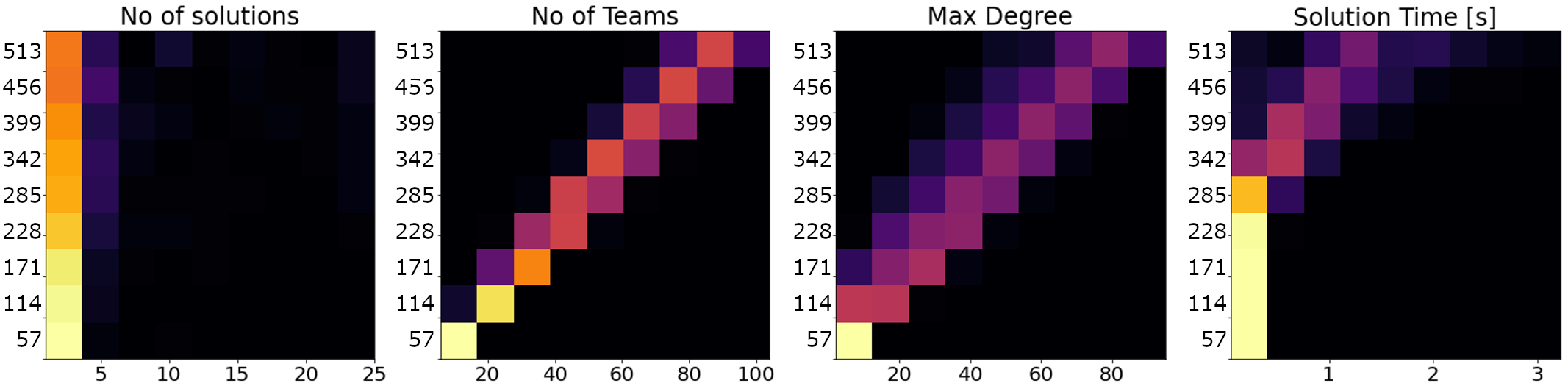}
\hfill
\includegraphics[width=0.39\textwidth]{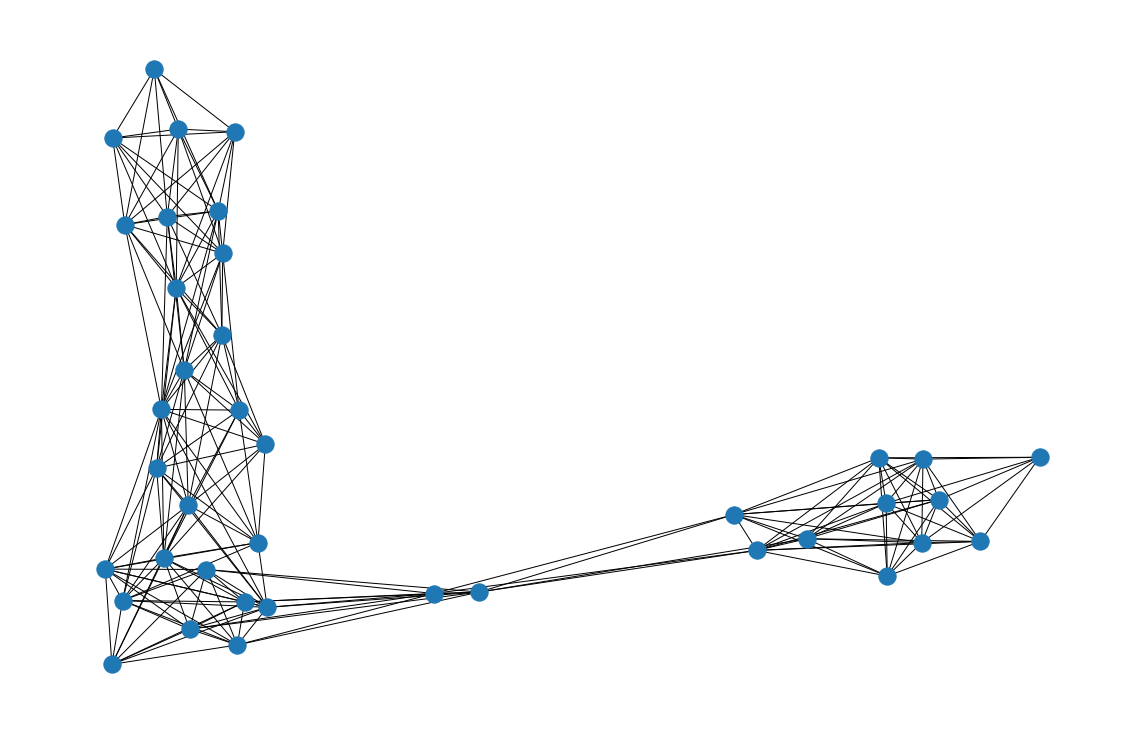}
\includegraphics[width=0.39\textwidth]{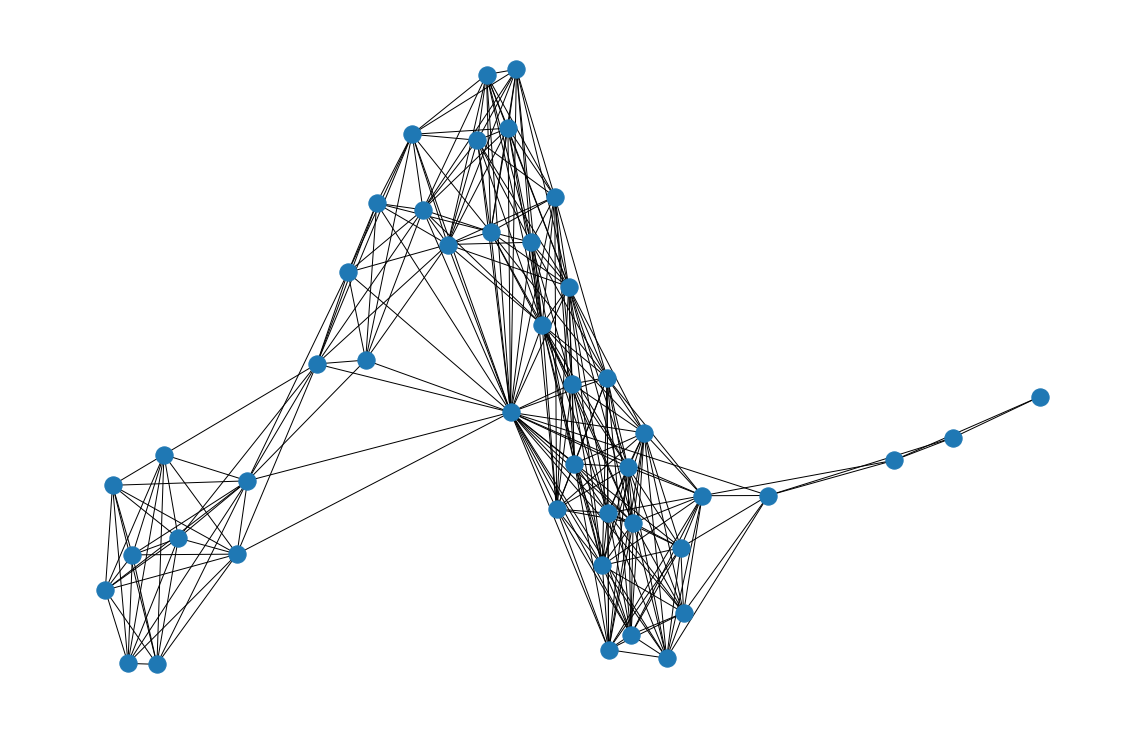}
\hfill
\caption{
{\bf Top:} Sample statistics for the Maximum Value Vertex Cover problem. The y-axis is number of beds in a scaled problem, with the full 513 corresponding to the real campus. Each line of boxes represents a normalised histogram, with the shading intensity indicating probability in a given box. The number of lowest-energy solutions grows only slightly with problem size, meaning the relative fraction of ground-state solutions decreases exponentially. The maximum degree of the collision graphs grows linearly with problem size. The runtime of a classical COIN-OR solver scales with the maximum graph degree. {\bf Bottom:} Two example collision graphs laid out with the spring method, showing a quasi-1D structure. The edges are present for reservations with overlapping requested dates. The position of vertices along the elongated direction comes directly from the dates requested by the vertex - reservation.}
\label{figSample}
\end{figure*}

We investigated the problem of room scheduling at an Australian Institute of Sport (AIS) campus facility\cite{AIS}, with the goal of accommodating the maximum number of incoming sports camp booking requests. Each request consists of a required number of beds and a duration. The number and sizes of rooms are based on the Canberra campus structure as of 2025, while the sizes and durations of requests are drawn from a probability distribution modelled on three months of reservation data from the first quarter of 2024 \cite{SM}. We prepare 30-day reservation stream samples and scale the number of available rooms in the entire campus facility, see Fig. \ref{figSample}. 

A single room can only be assigned to one request (team) at any given date. The total number of assigned beds must meet or exceed the requested number. These two constraints must be satisfied for a request to be accepted. 

Let $x_i$ be a binary variable indicating the assignment of room $i$ to request $x$. Let $R_x$ be the number of requested beds, and let $s_i$ denote the size of room $i$. To fulfil a request, the following constraint must be satisfied:

\be
R_x\leq \sum_i s_i x_i
\ee
To ensure that no room is assigned to multiple requests on the same date, we impose the following constraint:
\be
1\geq \sum_x {x_i} \cdot \begin{cases}
    1, & \text{if $x$ requests date $d$},\\
    0, & \text{otherwise}.
  \end{cases}
\ee

An alternative formulation of the room assignment constraint is as a vertex cover problem, where vertices represent requests and edges exist if two requests overlap on the same date. A valid vertex cover ensures that no two assigned requests share an edge:
\be
\sum_{<i,j>}x_ix_j=0
\ee

Let $N$ be the number of requests, $M$ the number of rooms, and $D$ the number of considered dates. The problem consists of searching an $N \times M$-dimensional binary space for a solution that satisfies $N + D$ linear inequality constraints. This problem falls within the domain of Linear Programming, specifically Binary Integer Programming, which is commonly addressed using Branch and Cut solvers. To translate this into the language of quadratic binary problems, inequality constraints must be converted into equality constraints using auxiliary variables \cite{QUBO_book}:

\be
R_x+H_x = \sum_i s_ix_i
\ee
where $H_x$ is an integer in the range $[0, S_{\max} - 1]$, with $S_{\max}$ being the maximum room capacity. Using a binary representation of $H_x$ introduces an additional $\log_2(S_{\max})$ bits per request, leading to a total of $N \cdot [M + \log_2(S_{\max})]$ bits for the entire problem. The equality constraint is incorporated into the optimization function as a quadratic energy penalty:
q
\be
\left(\sum_i s_i x_i-R_x-H_x\right)^2
\ee

The expansion of this constraint introduces all possible quadratic terms $x_i x_j$. On an annealing device, this requires full connectivity between all $x_i$ variables. Embedding a clique of size $K$ on a quasi-2D graph necessitates chains of bits to represent each variable, with the required number of bits scaling quadratically in $K$. On the state-of-the-art D-Wave Zephyr hardware, the number of required physical qubits per team scales approximately as $\frac{1}{8}K^2 + K$ \cite{ZephyrTopology}. This increases the number of required qubits per team from $M$ to approximately $\frac{M^2}{8}$. Adding further constraints further increases the total number of qubits required. 

A new machine originally projected to be available in 2024 with 7000 qubits would be able to represent fully connected problems of no more then about 220 bits. As the size of the machine increases in the future the required annealing time for hard problems is expected to increase exponentially with the machine size due to the gap problem\cite{Rajak2022,Jrg2009EnergyGI,Hen2011ExponentialCO}. {Information about maximum usable annealing time is not as easy to find as the historical chip sizes.} 

The current generation of quantum annealers cannot represent the entire problem. A full representation would require the number of logical qubits to be equal to the number of rooms (70) times the number of teams (60, assuming 30 days of operation), resulting in approximately 4000 logical qubits. These logical qubits must be embedded onto a limited-connectivity quantum processing unit (QPU), where each logical qubit is represented by a chain of strongly coupled physical qubits. Due to limited connectivity, the total required number of physical qubits scales with the number of edges in the problem graph. A fully connected component of 70 bits (representing a single team) requires around 600 physical qubits. Thus, just encoding teams without their interconnections would demand approximately 36,000 qubits. While a tenfold increase in qubit count may be feasible within five years \cite{Ichikawa2023CurrentNO}, achieving meaningful computations will also require substantial improvements in qubit quality.

\subsection{Approximate solutions}
Given current quantum hardware limitations, we explore a hybrid approach in which a smaller portion of the problem is solved on a quantum annealer. While this approach does not guarantee a global minimum, it has a well-defined runtime. Here we define it and investigate its performance.

The problem formulation naturally leads to two greedy algorithms. The first approach, referred to as \textbf{Greedy}, assigns teams one at a time, selecting empty rooms with minimal logic, prioritising efficient integer partitions of the team size.

\begin{figure*}[!htp]
\includegraphics[width=0.59\textwidth]{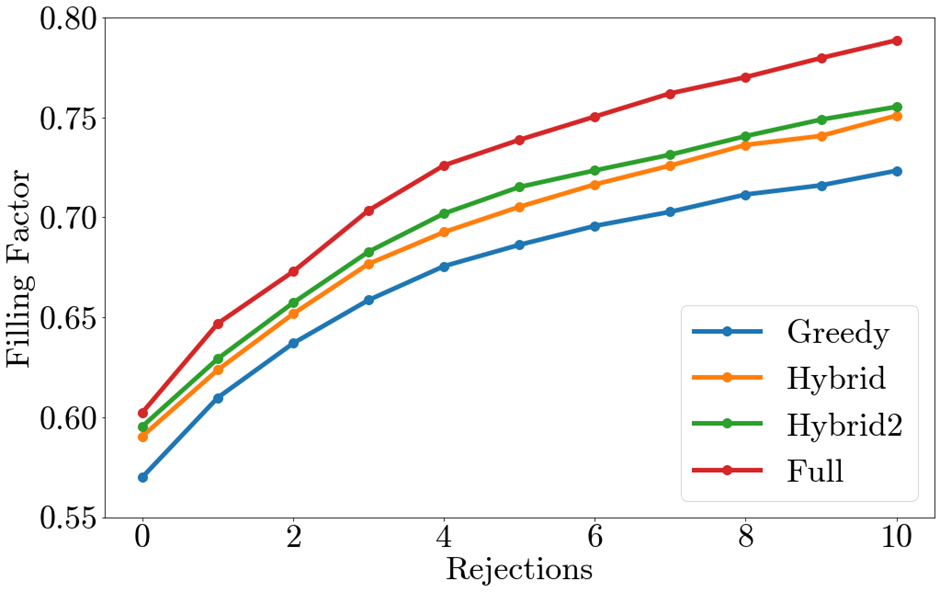}
\hfill
\caption{
Requests are drawn from a random distribution, and the filling factor is plotted at the first point where the solution fails to find a feasible rooming arrangement. The simulation then continues with this rejected request until the next failure, at which point another data point is recorded. Each point represents an average over 50 simulated reservation streams, which are kept identical for each method. The top solid red line represents the optimal solution of the full problem. Initially, the simulation runs for 30 days with a consistent solution method to establish the campus' initial state. The next 30 days serve as the test period, where solutions are sought using generated reservation streams while respecting room unavailability due to existing bookings, based on the prepared initial state. The results indicate that hybrid approaches---heuristic solutions combined with room-based optimisation---outperform simpler team-based methods.}
\label{fig1}
\end{figure*}

The second approach, \textbf{Hybrid}, prioritises satisfying the more complicated room constraint first. It assigns rooms one at a time, filling them with as many individuals as possible from the unassigned set. A heuristic value is assigned to each team to prioritise efficient room filling. At each step, the algorithm seeks to minimise the energy function:
\be
\sum_{<i,j>}x_ix_j-\sum_iE_ix_i
\ee
That problem is formally the \textit{Maximum Value Vertex Cover}, a well-known NP-hard problem. Solving this smaller problem is feasible with current QA devices. Important step here is preparing the problem for each iteration by assigning the values based on the reservation parameters:
\be
E_i=f(R,U_i,D_i,F_t)
\ee
where $R$ is the room size, $U_i$ is the number of unassigned individuals in team $i$, $D_i$ is the camp’s duration and $F_t$ represents the occupancy factor on date $t$. We test two functions referred to as \textbf{Hybrid 1} and \textbf{2}\cite{SM}. These are compared against the optimal and Greedy solutions (see Figure \ref{fig1}).
The function $f$ has been designed manually, but using machine learning to infer better value assignments would likely lead to improved performance. Combining QA and machine learning in this way is an interesting prospect.

\section{Quantum Annealing}

\subsection{Problem formulation for quantum annealer}

To evaluate the QA part of this problem independently instead of the using the Hybrid value assignments functions we instead assign the values randomly as follows:
\be
E_i=\frac{(\text{Duration})_i}{\text{Max Duration}}+\frac{\text{randInt}(-3,3)}{\text{Max Duration}},
\ee
where the random integer term helps prevent the generation of multiple degenerate ground-state solutions.

The quantum annealer used in this study is the D-Wave Advantage 2 prototype \cite{Advantage2}, which allows access to annealing times ranging from 5 ns to milliseconds for Ising problems without linear terms. Any Quadratic Unconstrained Binary Optimisation (QUBO) problem can be reformulated as an Ising problem, and any Ising problem can be expressed without on-site terms by introducing a single auxiliary spin variable.  

To illustrate this transformation, consider the variable substitution: $x_i\rightarrow \frac{s_i+1}{2}$, where $s_i$ is a spin variable taking values ${-1,1}$ (following a convention where spin \(1/2\) is scaled by a factor of 2). The vertex cover constraint then becomes:
\beas
\sum_{<i,j>}x_ix_j&=&\left(\sum_{<i,j>}\frac{s_is_j}{4}+\sum_{<i,j>}\frac{s_i+s_j}{4}\right)
\\&=&\left(\sum_{<i,j>}\frac{s_is_j}{4}+\sum_{i}\frac{O_is_i}{2}\right),
\eeas
where $O_i=\sum_{j\in<i,j>}1$ represents the number of vertices neighbouring $i$. 

This formulation highlights a challenge in quantum annealing: the on-site term scales linearly with the number of neighbors, requiring hardware capable of supporting a growing parameter range as the problem size increases.

To remove the linear terms, we introduce an auxiliary spin variable $X$:
\beas
\left(\sum_{<i,j>}\frac{s_is_j}{4}+\sum_{i}\frac{O_is_i}{2}\right)\rightarrow\left(\sum_{<i,j>}\frac{s_is_j}{4}+\sum_{i}\frac{O_is_iX}{2}\right).
\eeas
This transformation ensures that the original solutions ${s_i}$, remain valid when $X_i=1$, while introducing a degenerate set of solutions ${-s_i}$, when $X_i=-1$. An additional benefit of this approach is that flipping the $X$ spin effectively flips all other spins, acting as a global XOR operation that reduces the maximum Hamming distance between configurations by half.

To handle large values of $O_i$, we can replace $X$ with a more complex structure:  
\beas
\sum_{i}\frac{O_is_iX}{2}\rightarrow\sum_{k=1}^M\sum_{i}\frac{O_is_iX_k}{2M}-J\sum_{k,l=1}^MX_kX_l.
\eeas
Here $M$ auxiliary spins $X_k$ are introduced, each contributing $\frac{1}{M}$ of the required weight, thereby reducing the parameter range. The second term enforces strong coupling among the \(X_k\) variables, ensuring they remain aligned. To preserve the problem’s low-energy structure, we set \(J(M-1) = 1\). Exceeding this value introduces higher-order errors, which are discussed in Supplementary Material \cite{SM}.  
 
The final step is formulating the Ising Hamiltonian. Since the cover condition $x_ix_j=0$ is always satisfied in solutions, the monomial $x_i$ is equivalent to $x_i(1-x_j)$. This allows us to express the energy function as:  
\be
\sum_iE_ix_i\stackrel{sol}{=}\sum_{<i,j>}w_{ij}E_ix_i(1-x_j), \sum_i w_{ij}=1
\ee 
with the latter formulation allowing for higher values of $E_i$ to be used without breaking the edge exclusivity condition $x_ix_j$. It the first case $x_ix_j$ has a unit energy difference between allowed and broken states and would be broken, when $\min(E_i,E_j)>-1$. In the latter case, $x_i x_j$ can be arbitrarily high for two bits without changing the energy gap between allowed and forbidden solutions. For highly connected bits it approaches the first limit for the choice $w_{ij}=\frac{1}{O_i}$ used in this paper. Weights $w_i$ could be further optimised to strengthen the connections more likely to be broken, improving in practice over the standard formulation in \cite{Lucas2014}.

Finally, the chain strength must be calibrated. This parameter determines the energy cost of breaking a chain of qubits, which should behave as a single logical qubit after embedding onto the quantum processing unit (QPU). These chains (trees) are created by minor-miner\cite{Cai2014APH} embedding. The optimal chain strength is the smallest value that preserves the chains in solutions, which is determined by calculating the maximum possible energy difference when breaking a chain. Setting the chain strength too high can overshadow problem couplings, degrading performance.

\subsection{Quantum Annealer Calibration}

Here we describe all the steps performed in order to obtain best results from QA, in particular our new multi-qubit, problem-informed calibration scheme.

We first start with a standard calibration technique \cite{Chern2023TutorialCR} initialising all Ising weights to zero, then generating a sample of chain states from the annealer, while adjusting flux-bias offsets \cite{Harris2009CompoundJC} to achieve zero average magnetisation for each chain. 

Then we consider a partial graph cover problem. That means generating a set of covers - selected vertices with a condition of no edge existing between two selected vertices $\sum_{<i,j>}x_ix_j=0$. Crucially we can generate solutions of this problem very quickly, as removing the value terms $V_i$ makes the problem more simple.
To generate an unbiased distribution of partial graph covers, we employ a Monte Carlo method:


\begin{algorithm}
\caption{Monte Carlo Vertex Cover Sampling}
\begin{algorithmic}[1]
\State \textbf{Input:} Graph $G=(V,E)$, number of samples $S$
\State \textbf{Output:} Vertex cover statistics
\State Initialize $\textit{Cover Statistics}$
\State Compute adjacency list \textit{neighbours}
\For{$s = 1$ to $S$}
    \State Initialize $\textit{current\_cover} \gets \emptyset$, $\textit{possible\_choices} \gets V$
    \State Set $\textit{Trajectory\ Probability} \gets 1$
    \While{$\textit{possible\_choices} \neq \emptyset$}
        \State Randomly select $v \in \textit{possible\_choices}$
        \State Add $v$ to $\textit{current\_cover}$
        \State Update $\textit{Trajectory\ Probability}=\textit{Trajectory\ Probability}\times\#(\textit{possible\_choices})\times\#(\textit{current\_cover})$
        \State \textbf{Gather statistics:} Update $\textit{Cover Statistics}$ with weight $\textit{Trajectory\ Probability}$
        \State Remove $v$ and its neighbours from $\textit{possible\_choices}$
        \State Increment $\textit{Total\ Covers}+=\textit{Trajectory\ Probability}$
    \EndWhile
\EndFor
\State Normalize $\textit{Cover Statistics}$ dividing by $\textit{Total\ Covers}$
\State \textbf{Return} $\textit{Cover Statistics}$
\end{algorithmic}
\end{algorithm}
The problem is run on Quantum Annealer and from the sample of results we generate occurrence statistics for each pair of bits $(x_i,x_j)$. The Monte-Carlo samples statistics $P^{(00,01,10,11)}_{i,j}$ are compared to QA sample statistics $Q^{a,b}_{i,j}$. As within a single QA sample results may be auto-correlated - some local magnetic fields persists between subsequent annealing processes - we instead use an average from multiple samples run with much larger time separation. Comparing the obtained statistics to desired statistics gives corrections to be added to the QUBO model:
\beas
&\epsilon&\Delta^{00}\text{erf}(|\Delta^{00}|/\sigma)(1-x_i)(1-x_j),\\
&\epsilon&\Delta^{10}\text{erf}(|\Delta^{10}|/\sigma)(x_i)(1-x_j),\\
&\epsilon&\Delta^{01}\text{erf}(|\Delta^{01}|/\sigma)(1-x_i)(x_j),\\
\text{where} &&\Delta^{ab}=(Q^{ab}_{ij}-P^{ab}_{ij})),\\
\eeas
with $\sigma$ being the variance of errors $\Delta$ for a single shot. $\sigma$ is estimated on the first samples, before applying corrections. The scale of the corrections $\epsilon$ decays exponentially with iterations and they are scaled by the $\text{erf}$ function to account for the random fluctuations. The iteration is stopped once average $|\Delta|$ is smaller then $\frac{\sqrt{2}}{\pi}\sigma\approx 1.1 \sigma$ - the expected value of difference of random variables $E(|x-y|)$. Corrections are applied to only three binomials, as the last possible QUBO term $(x_i)(x_j)$ is linearly dependent on the rest by the probability normalisation condition. 

\begin{figure*}[!tp]
\includegraphics[width=0.32\textwidth]{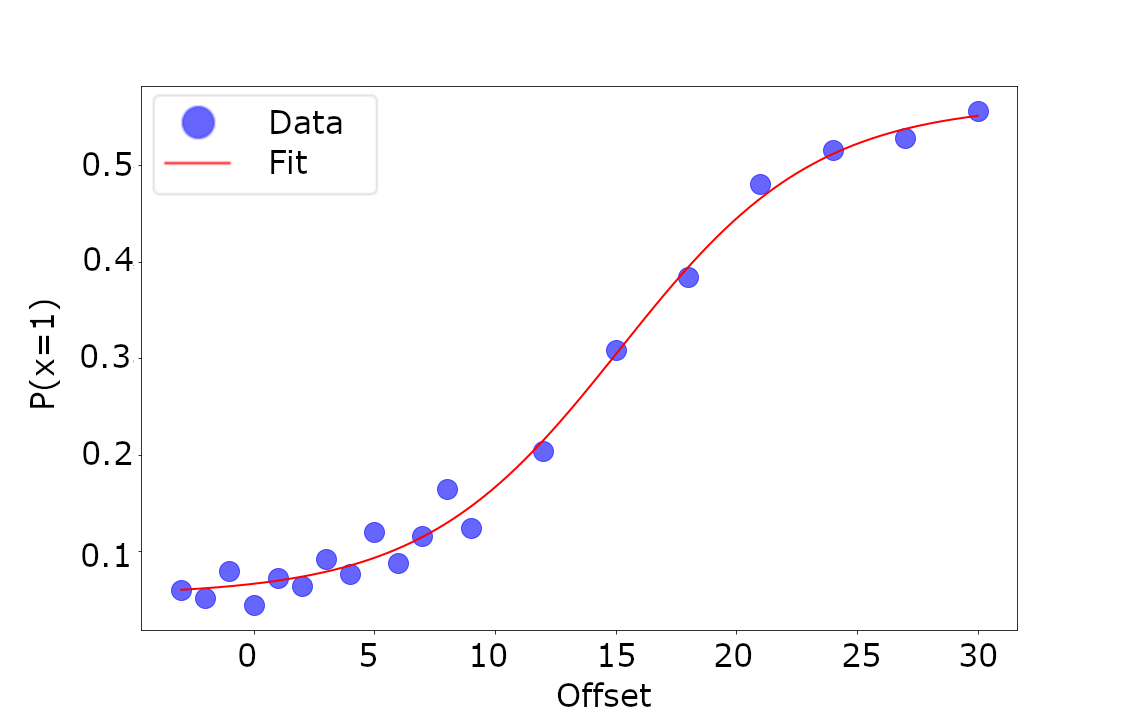}
\includegraphics[width=0.32\textwidth]{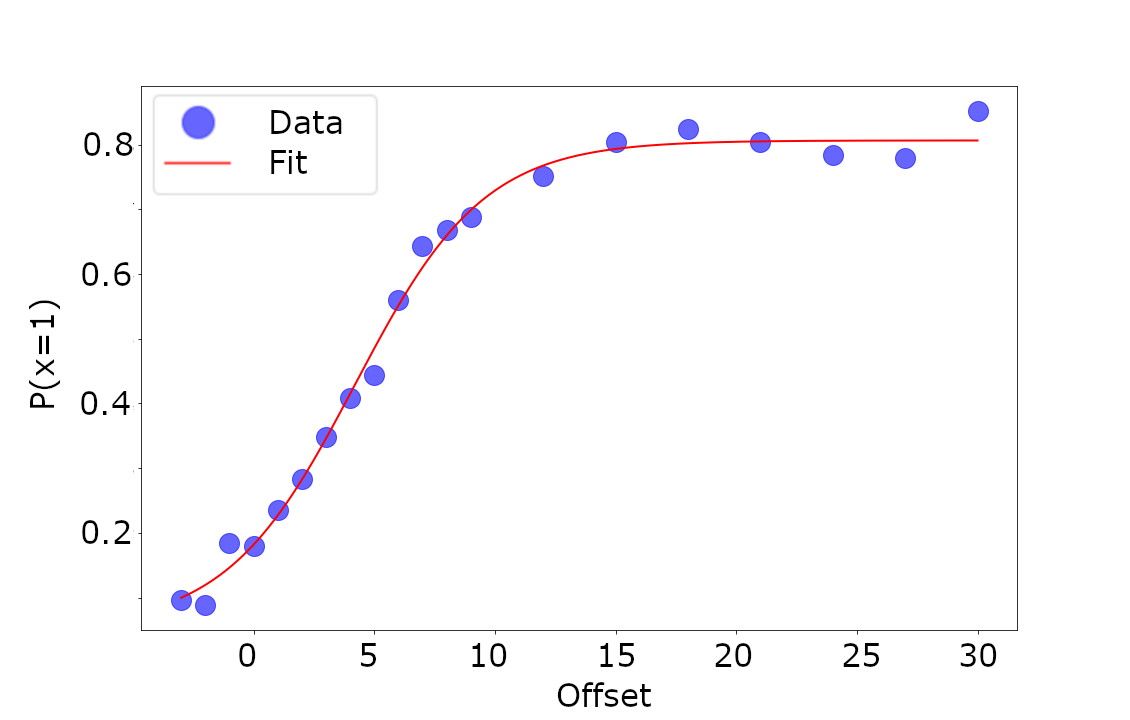}
\includegraphics[width=0.32\textwidth]{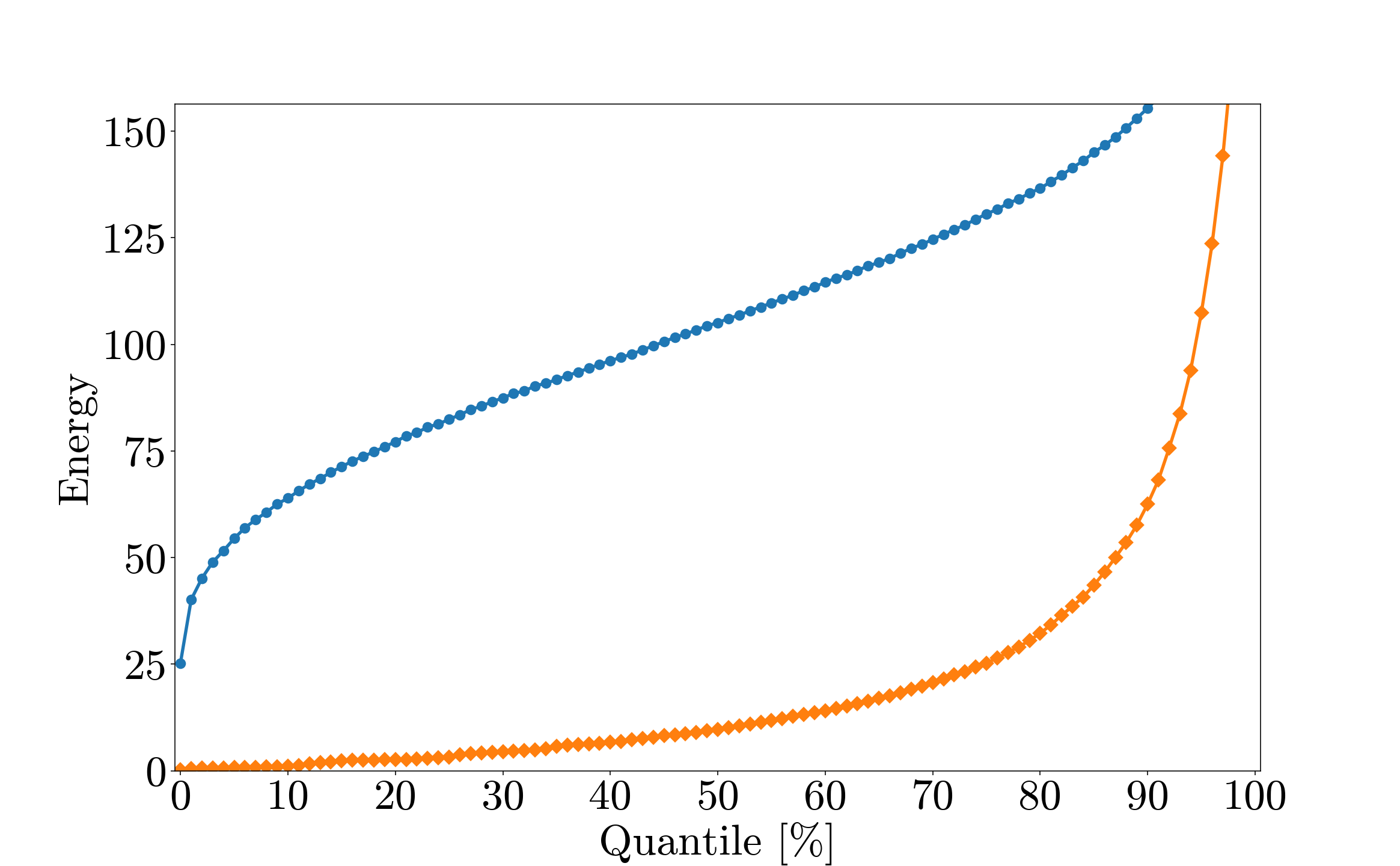}
\hfill
\caption{
{\bf Left and centre:} Two example bits being calibrated to find the offset strength scaling. The calibration is started with zero weighted problem. The offset is introduced only on one logical bit, which should make it appear in all selected covers. In reality this transition is broad and its width varies 	strongly. Measured statistics are fitted with sigmoid function: $w$, $b+\frac{(a-b)}{2}\cdot\left[\text{tanh}(\frac{V-V_0}{w})+1\right]$. The intended strengths are scaled by triple the fitted width $w$, as $\text{tanh}(1.5)+1$ corresponds to about $95\%$ of the total height. {Right:} Comparison of energy statistics before and after calibration. 
}
\label{fig2}
\end{figure*}

In the last step we calibrate individual variable strengths by applying a negative offset $V_i$ to a single qubit and measuring its inclusion probability in the final solution. In the perfect case any offset leads to 100\% inclusion of the bit in generated covers. In reality this transition from follows a sigmoidal profile, fitted as:
\be
b+\frac{(a-b)}{2}\left[\text{tanh}\left(\frac{V-V_0}{w}\right)+1\right].
\ee
This transition happens with a finite width and for individual $i$ it may be very disparate, as shown in Fig. \ref{fig2}. The width of this transition is used to scale the values $V_i$. 

This new calibration setup allows for much more in-depth calibration. We encoded almost all problem parameters, with the exception of the value clauses and hence can address imperfections present in almost exactly the configuration we will want to solve in the hard Vertex Cover problem. Crucially, as the Hybrid algorithm progresses, the collision graph stays the same and the only parts which changes are the values $V_i$, making this calibration universal.

\section{Results and Outlook}
\subsection{Quantum Annealing}
\begin{figure*}[!tp]
\includegraphics[width=0.32\textwidth]{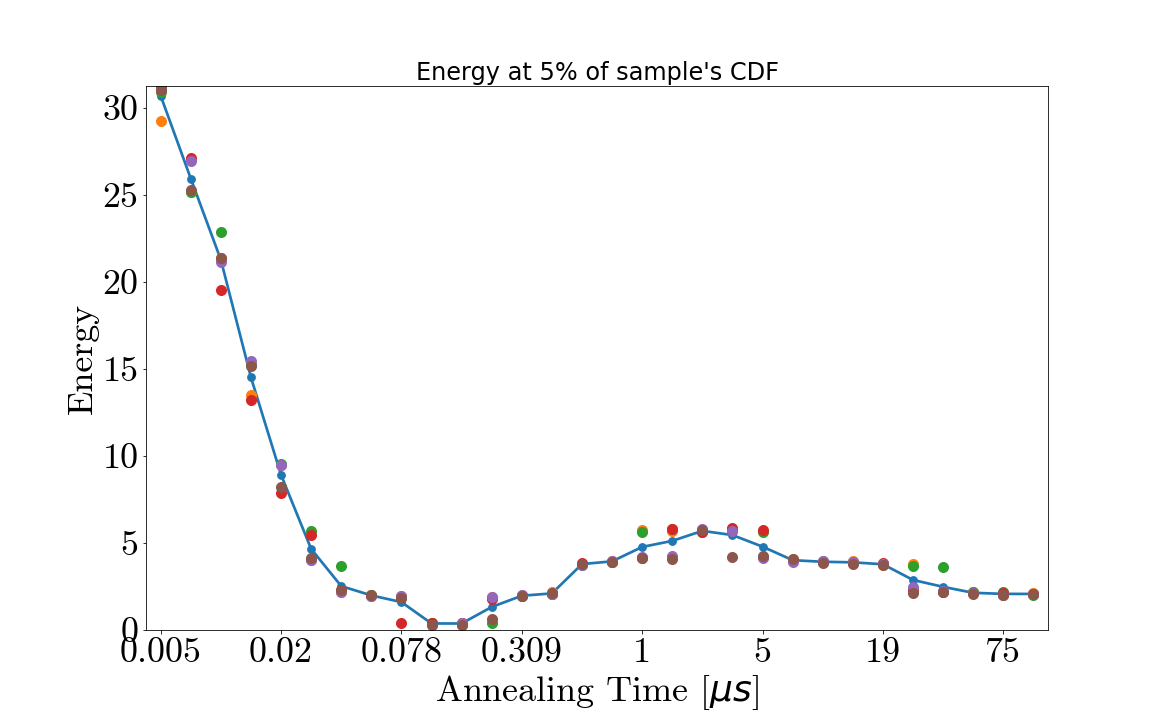}
\includegraphics[width=0.32\textwidth]{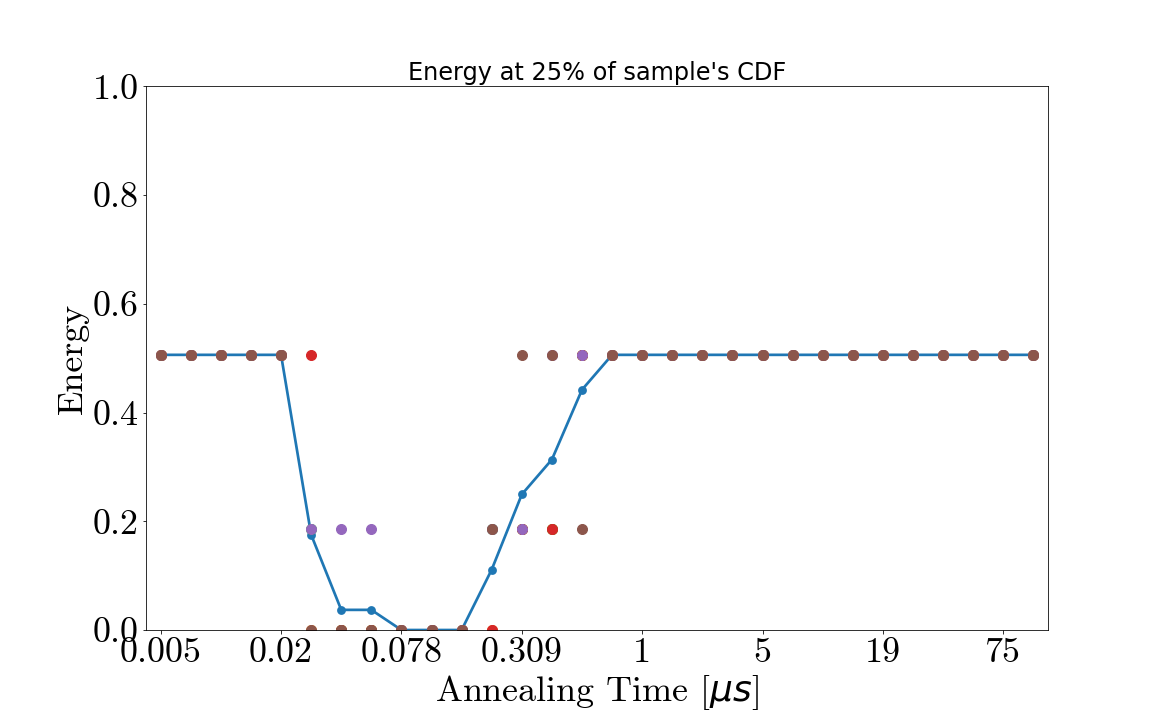}
\includegraphics[width=0.34\textwidth]{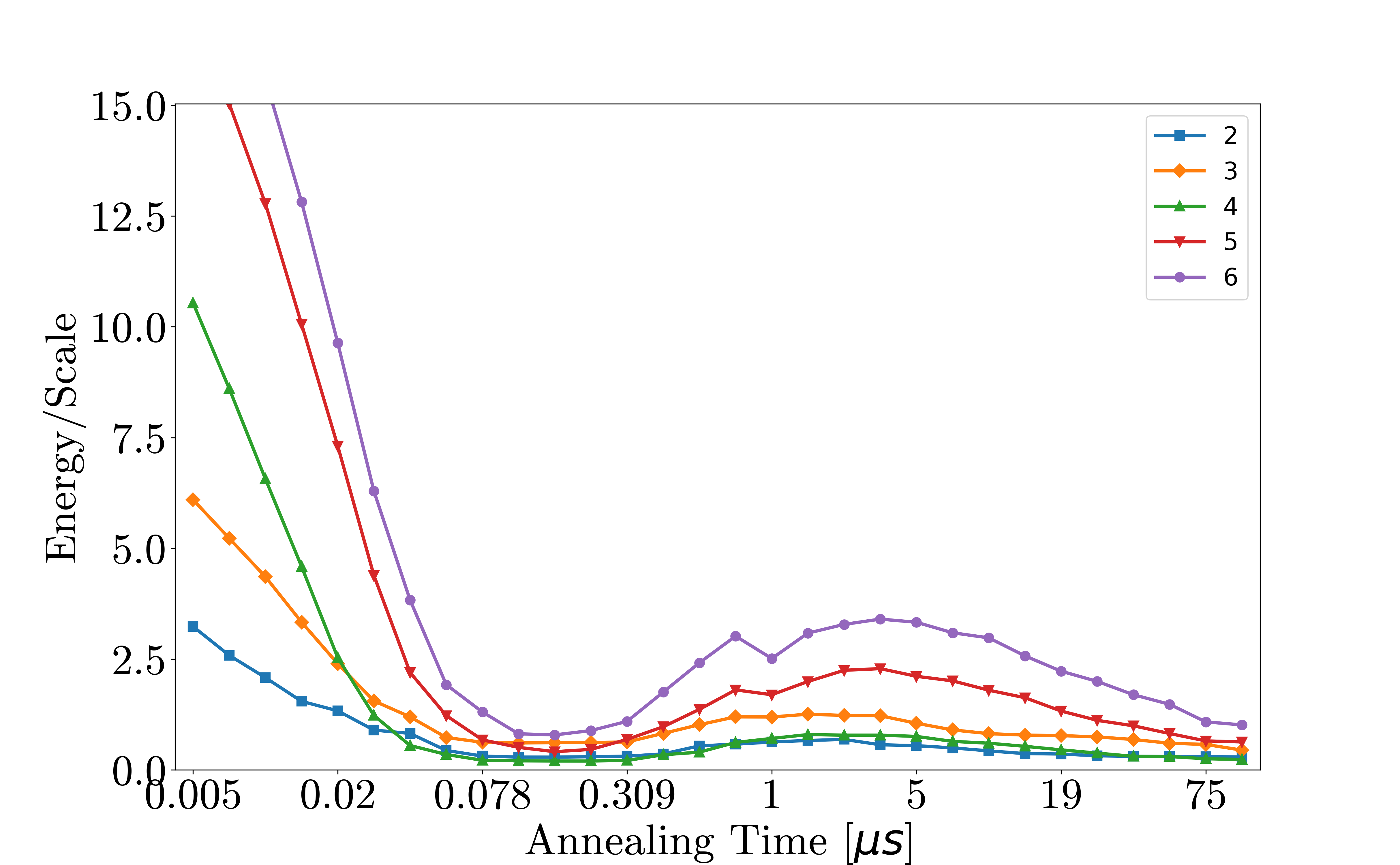}

\caption{{\bf Left and centre:} Energy dependence of raw samples obtained from the quantum annealer and post-processed samples refined using a steepest descent algorithm. The sample scaling factor size is 2, corresponding to 35 logical bits (41 including the X - XOR bits). The left plot shows energy at the $5\%$ quantile, while the centre plot shows the $25\%$ quantile. Each of the points represents an average of a sample of 1000 quantum annealing runs.
{\bf Right:} The average energy at the $25\%$ quantile dependence on annealing time. Each line is an average over 7 random realisations of the problem. Different series correspond to different sample sizes (see legend). The scaling factor $s$ sets the number of beds in the campus to $57s$. Energy is scaled by the scaling factor - in unscaled units the linear growth in energy corresponds to roughly exponential growth of the number of better solutions available below this energy.
}
\label{figEne}
\end{figure*}

The primary conclusion of this study is that the optimal annealing time for this optimisation problem coincides with the quantum coherence time, which is approximately $100~\text{ns}$. This is evident when analysing the average energy across different quantiles of large annealing result samples (see Fig. \ref{figEne}). Even after post-processing these samples with a classical approximate solver, this conclusion remains valid, albeit with a slightly reduced effect. Notably, annealing for durations exceeding the coherence time leads to a decline in sample quality. Only after an extended period does the sample quality begin to improve again, likely due to thermal annealing effects. It takes approximately $100~\mu\text{s}$ for thermal annealing to restore sample quality to the level achieved with rapid annealing. This timescale is comparable to that of classical local neighbourhood search solvers, which typically take a few microseconds per sample.

Our findings align with observations made a decade ago in \cite{PhysRevA.92.052323}, where an optimal annealing time in the range of tens of nanoseconds was reported. This suggests that, despite advances in hardware, the fundamental timescale for effective annealing has not significantly changed.

This minimum annealing time remains consistent across all investigated problem sizes. However, as the problem size increases, the average sample energy rises significantly. The performance of quantum annealing is relatively poor for this problem, likely due to the energy scales involved. While the individual coupling strengths were optimised, the energy contribution from local magnetisation increases linearly with the number of connections. Additionally, the transition widths observed during the offset calibration procedure tend to grow with problem size. This suggests that in larger, more densely connected graphs, longer-range couplings degrade performance by making the problem more delocalised. 

\subsection{Calibration}
A key contribution of this work is the new calibration scheme, which significantly improves annealing results by correcting biases in qubit interactions in addition to standard single qubit calibration. Our approach to problem formulation refines energy scales, reducing errors caused by problem size growth and long-range couplings. Future research should further optimise this calibration and explore embedding strategies to mitigate hardware limitations.

\subsection{Hybrid Methods}
While quantum annealing struggles with highly connected graphs, hybrid quantum-classical methods and improved calibration may extend its applicability. Identifying problem classes where quantum methods yield some type of performance gain over classical solvers remains a critical next step. While there is no unified theorem saying the more you solve exactly, the better greedy algorithm becomes, in this problem it is the case. It would be interesting to be able to solve a larger portion of the problem, likely moving closer to optimal solution still, but for now we have saturated, what the QA machine can achieve.

{\bf Acknowledgements}
This research was supported by funding from the CSIRO Quantum Technologies Future Science Platform. We would like to acknowledge valuable discussions with Peter Tyson and Clement Chu from CSIRO, as well as Michael Hall and Jim Gall from NEC Australia, whose insights contributed to the development of this work. We also thank the Australian Institute of Sport for providing access to their reservation data and for proposing the optimisation problem that motivated this study.


\clearpage
\section*{Supplementary Material}

\subsection*{Random sample details}

The Australian Institute of Sport (AIS) campus facility\cite{AIS} booking requests are drawn from a probability distribution modelled on three months of reservation data from the first quarter of 2024 \cite{SM}. The reservation size distribution was fitted to data as Gamma probability distribution:
\beas
\frac{1}{\Gamma(\alpha)\theta^\alpha}x^{\alpha-1}\exp{-x/\theta}
\eeas
$x$ is the variable, $\alpha=5.86$ and $\theta=5.72$. This corresponds to the mean $\mu=33.6$ and variance $\sigma^2=197$. The random durations were drawn directly based on the histogram of the sample provided. This data had peaks around specific times like 3 day 5 days and some longer durations charactersic of sports training camps.

\subsection*{Hybrid Value Function}

In the \textit{Maximum Value Vertex Cover} the values based on the reservation parameters:
\beas
E_i=f(R,U_i,D_i,F_t)
\eeas
where $R$ is the room size, $U_i$ is the number of unassigned individuals in team $i$, $D_i$ is the camp’s duration and $F_t$ represents the occupancy factor on date $t$. We

 A heuristic value assigned to each team in the Hybrid 1 approach was designed to prioritise efficient room filling:
\beas
V_i=min(R,U_i)^\alpha\cdot D_i,
\eeas
where $\alpha=2$ controls the preference for fully filling a room.

The Hybrid 2 approach prioritises assignments based on expected occupancy levels, favouring assigning higher value to teams that overlap with high-in-demand dates:
\beas
V_i=min(R,U_i)^2\cdot D_i \cdot [\text{max}({F_t})]^3,
\eeas
where $F_t$ represents the occupancy factor on date $t$.

\subsection*{Results of calibration}
This section contains additional cumulative density function of energy from quantum annealing samples obtained before and after calibration procedure for different samples. \\
\includegraphics[width=0.45\columnwidth]{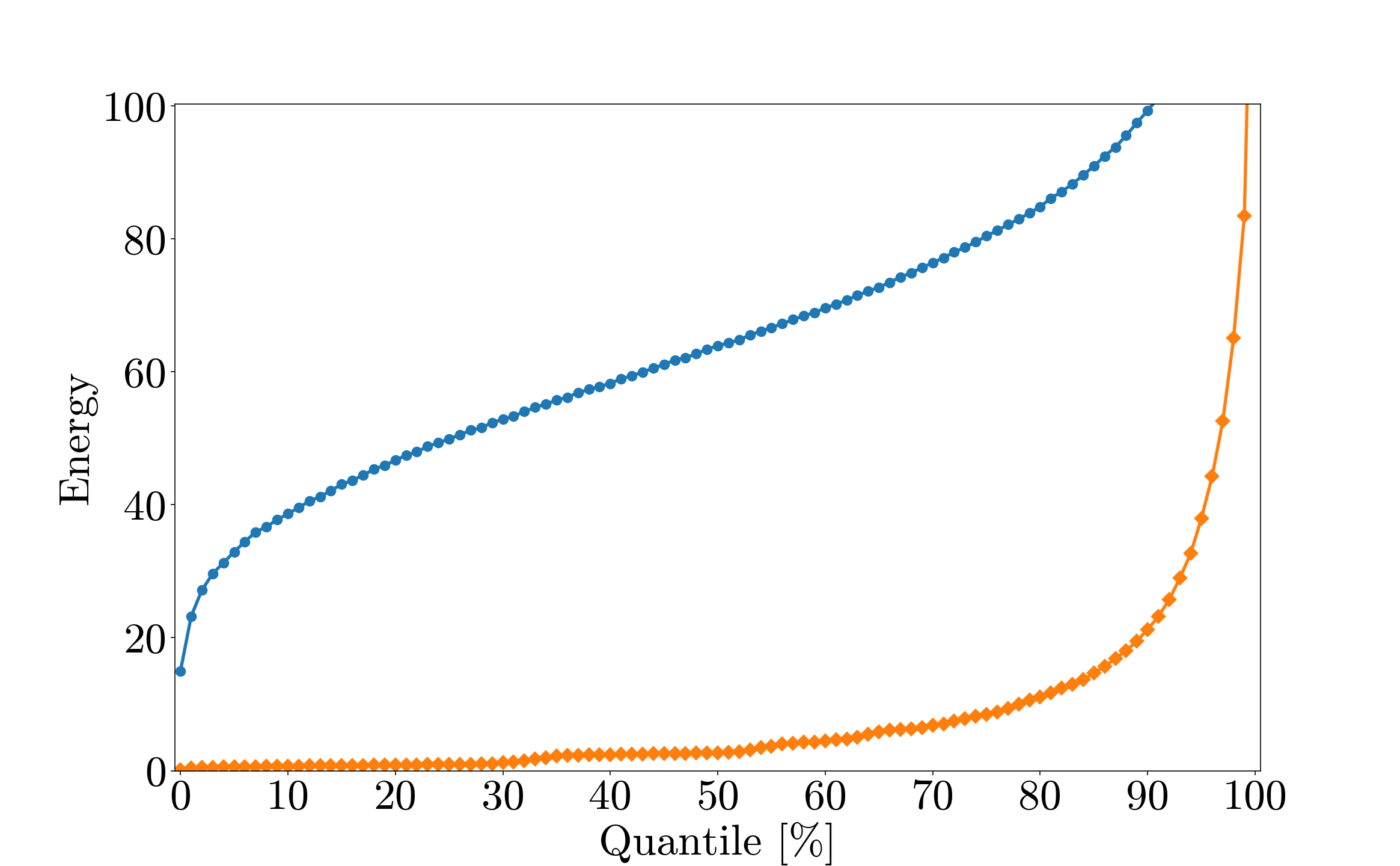}
\includegraphics[width=0.45\columnwidth]{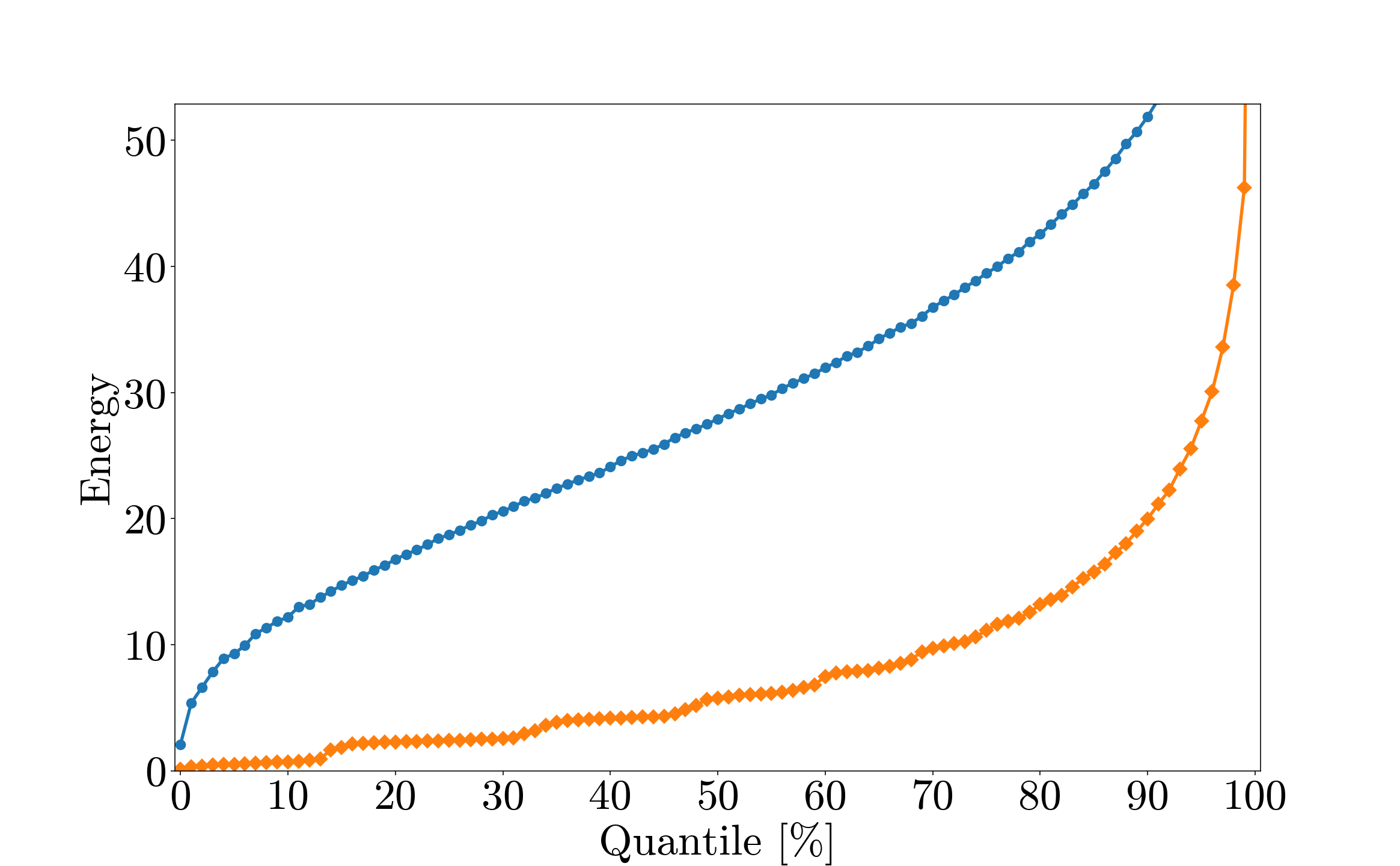}\\
\includegraphics[width=0.45\columnwidth]{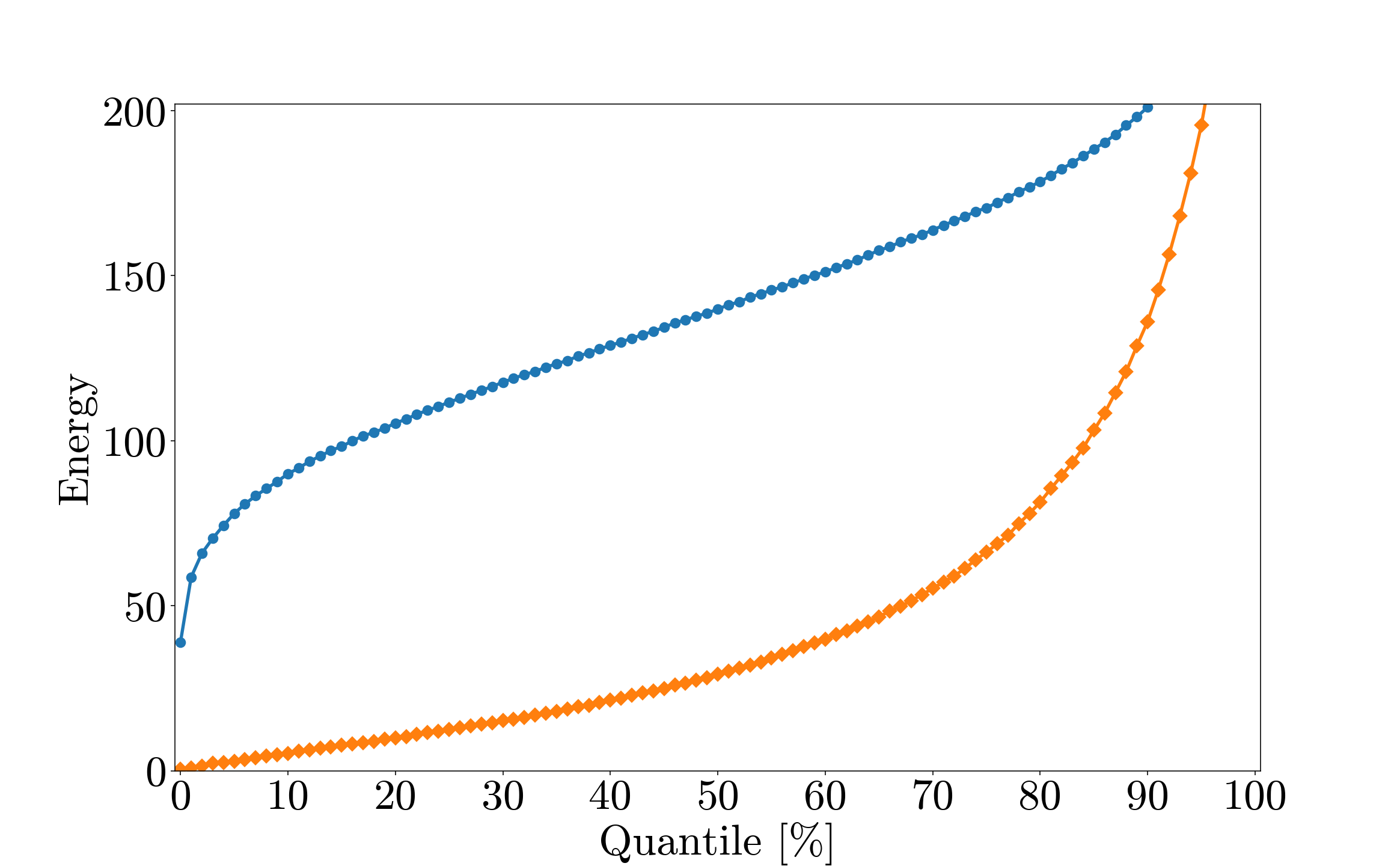}
\includegraphics[width=0.45\columnwidth]{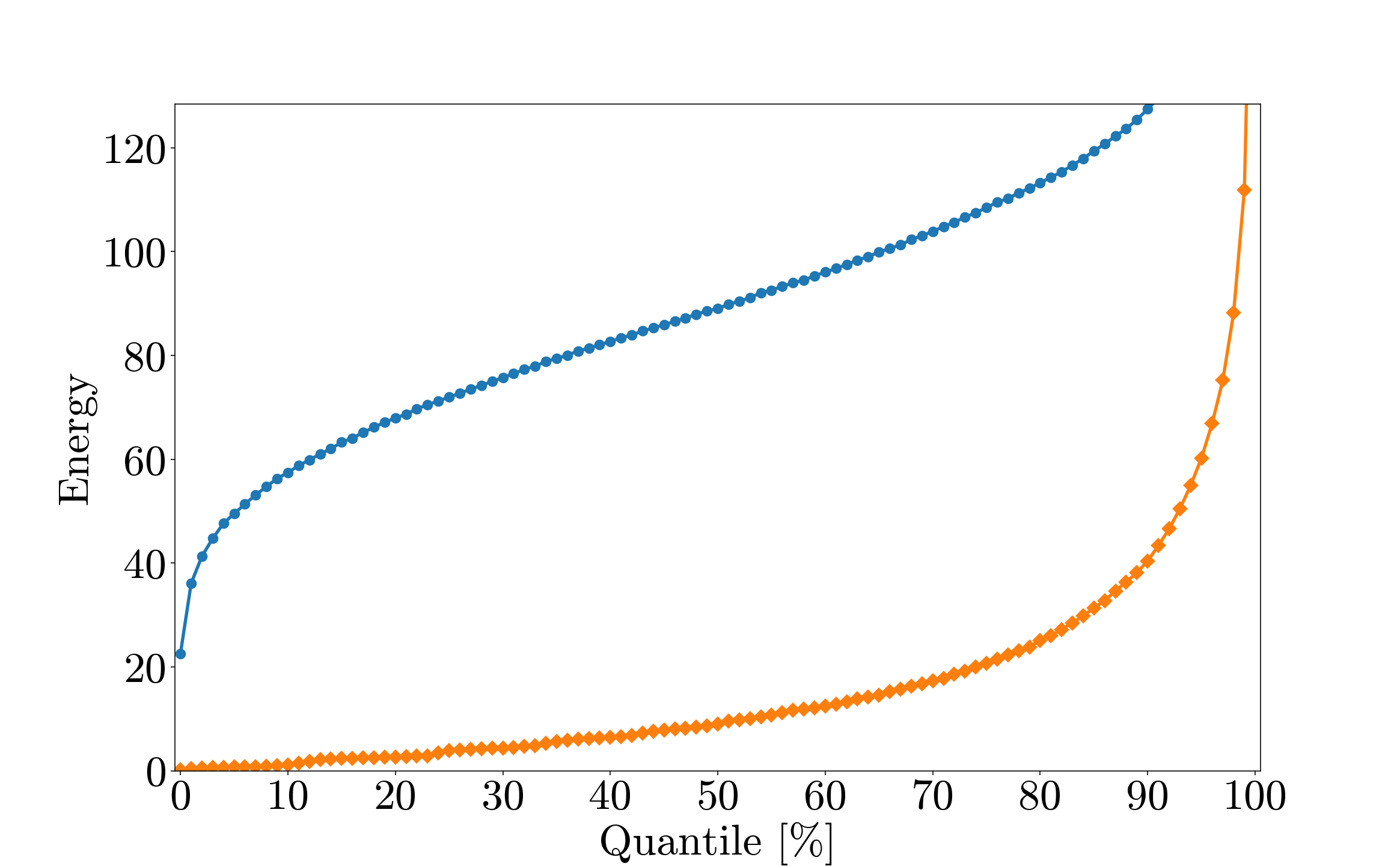}\\
\includegraphics[width=0.45\columnwidth]{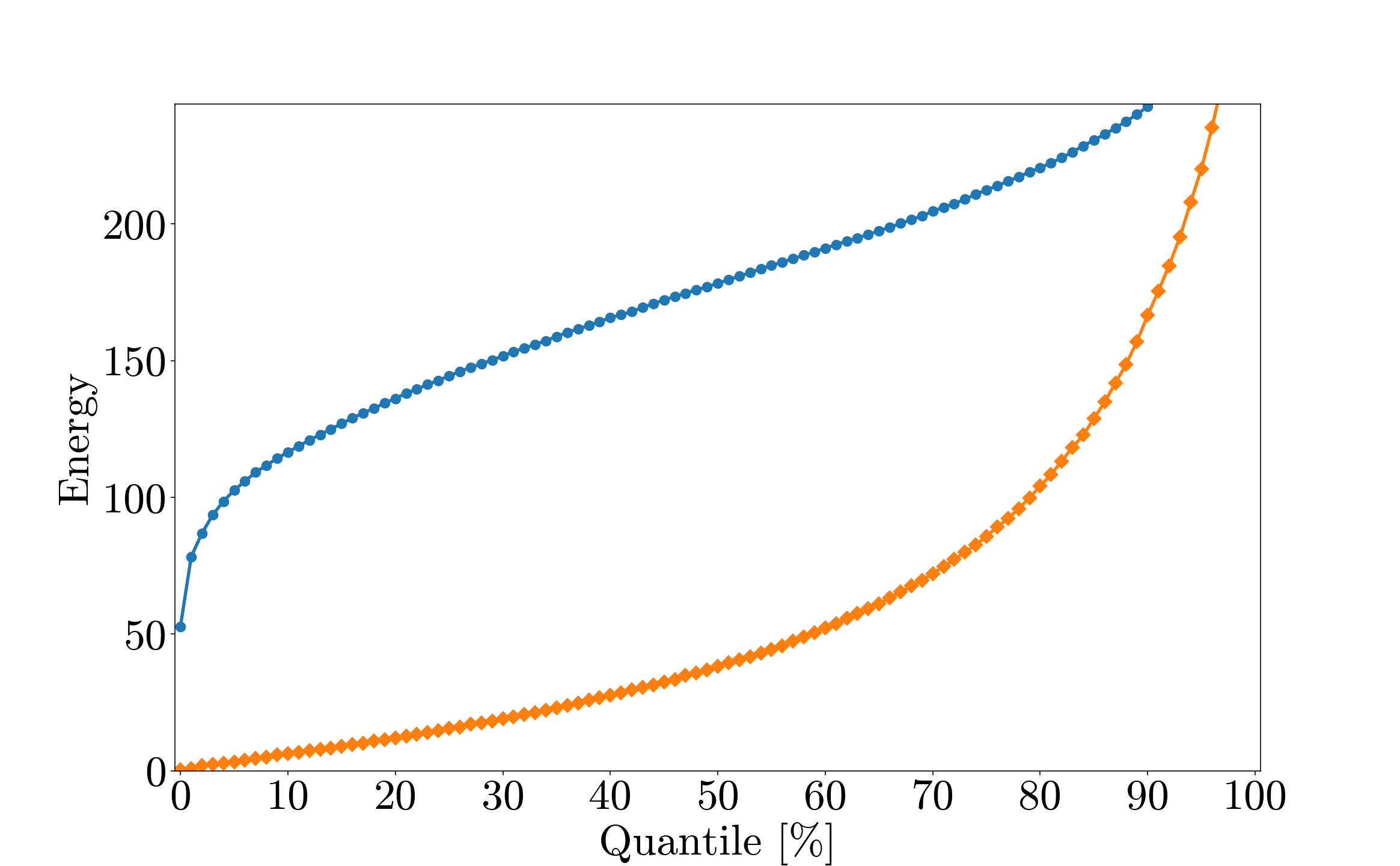}
\includegraphics[width=0.45\columnwidth]{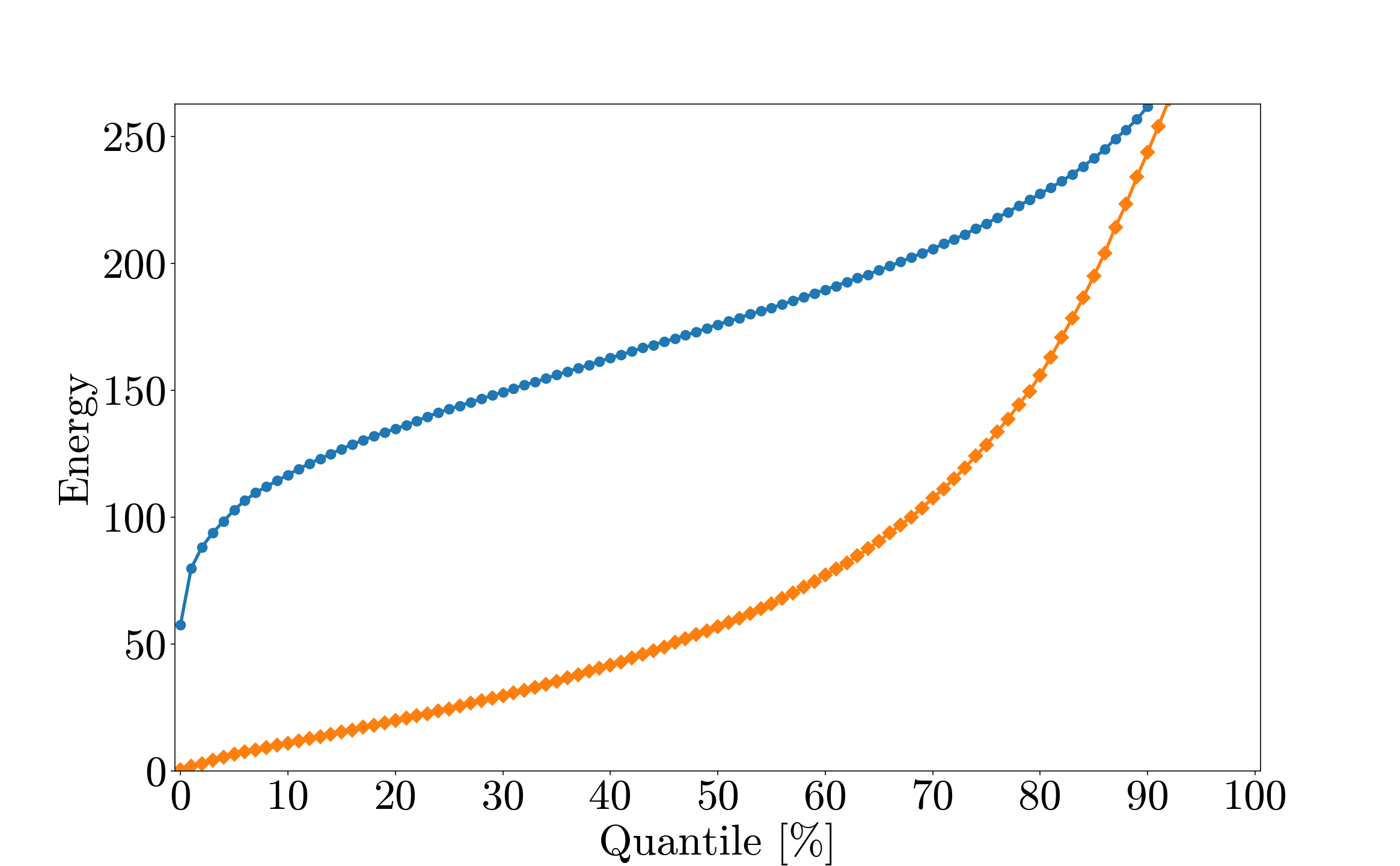}\\

\subsection*{Additional plots --- annealing time}
Below additional plots of energy dependence on annealing time for different quantiles of the sample are presented:\\
\includegraphics[width=0.45\columnwidth]{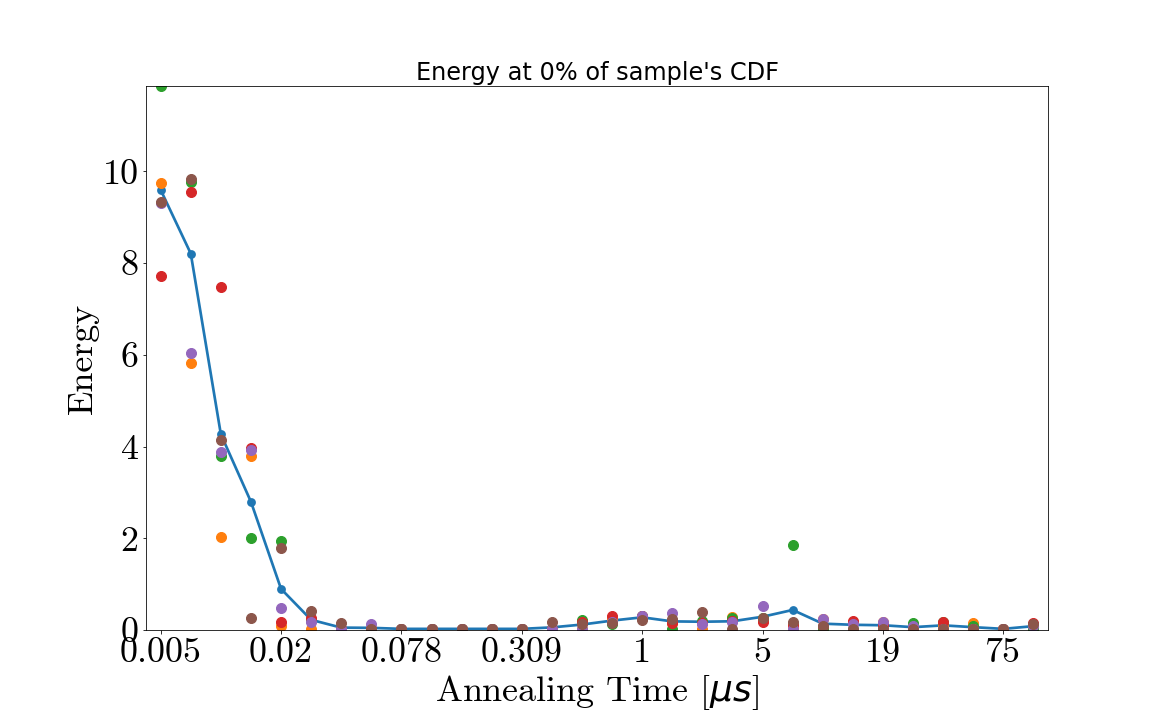}
\includegraphics[width=0.45\columnwidth]{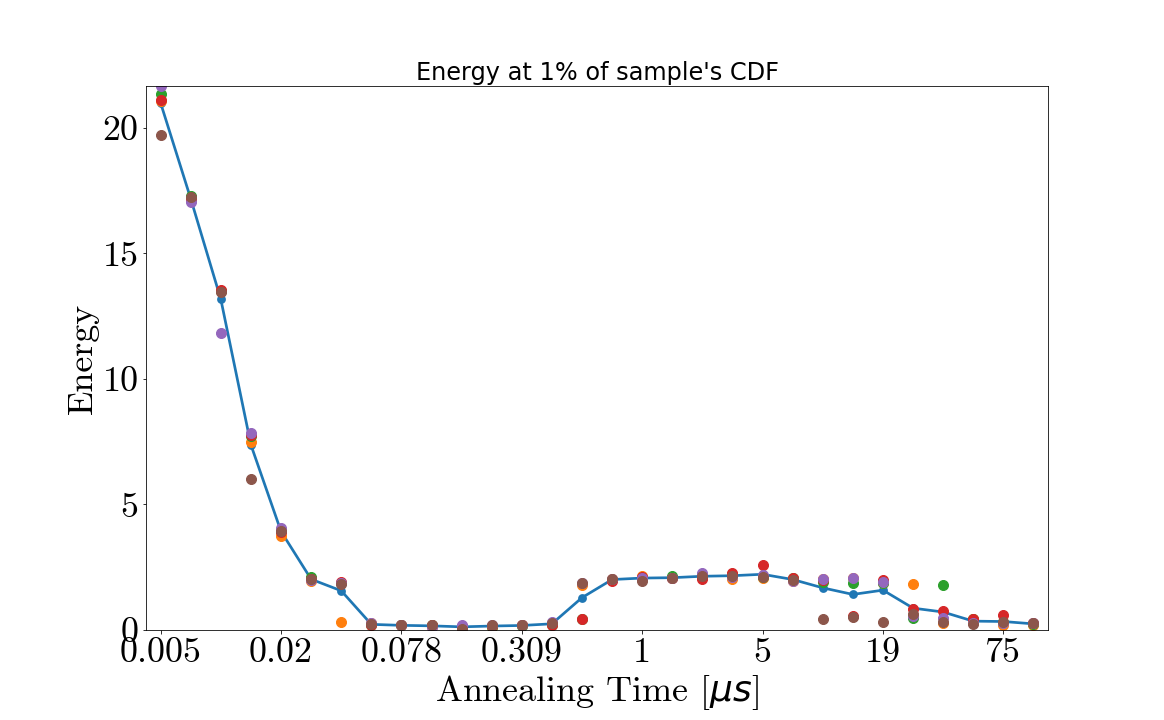}\\
\includegraphics[width=0.45\columnwidth]{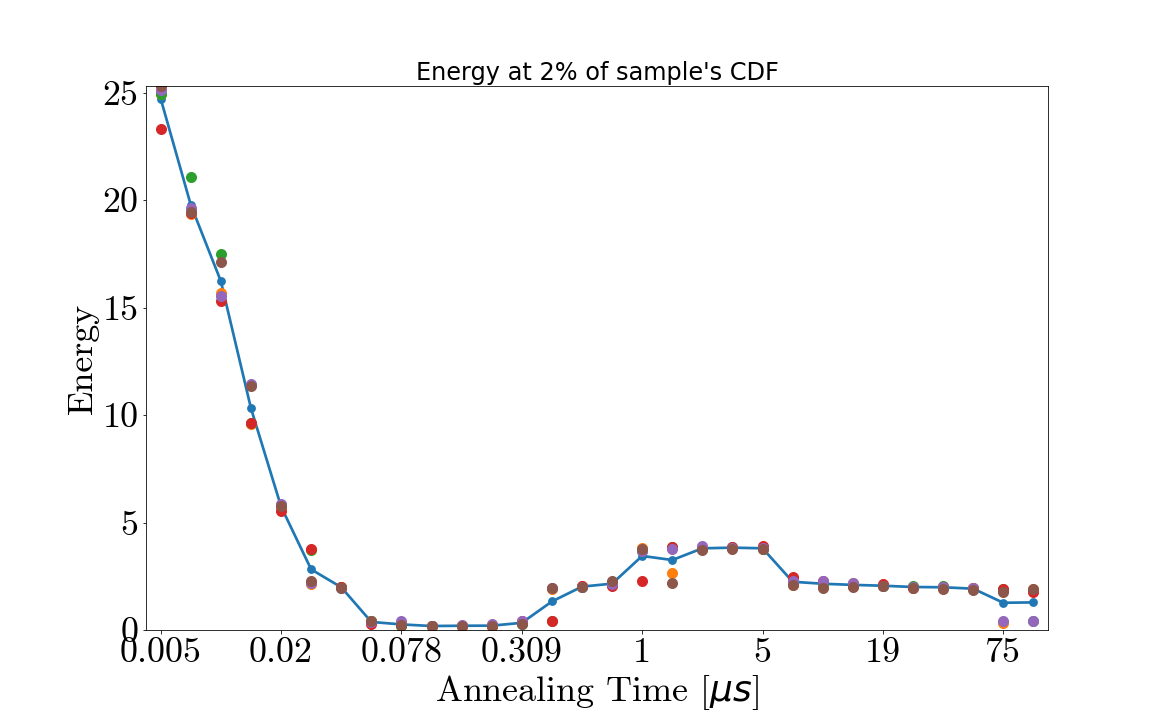}
\includegraphics[width=0.45\columnwidth]{HCAL/file28CDF0_5.png}\\
\includegraphics[width=0.45\columnwidth]{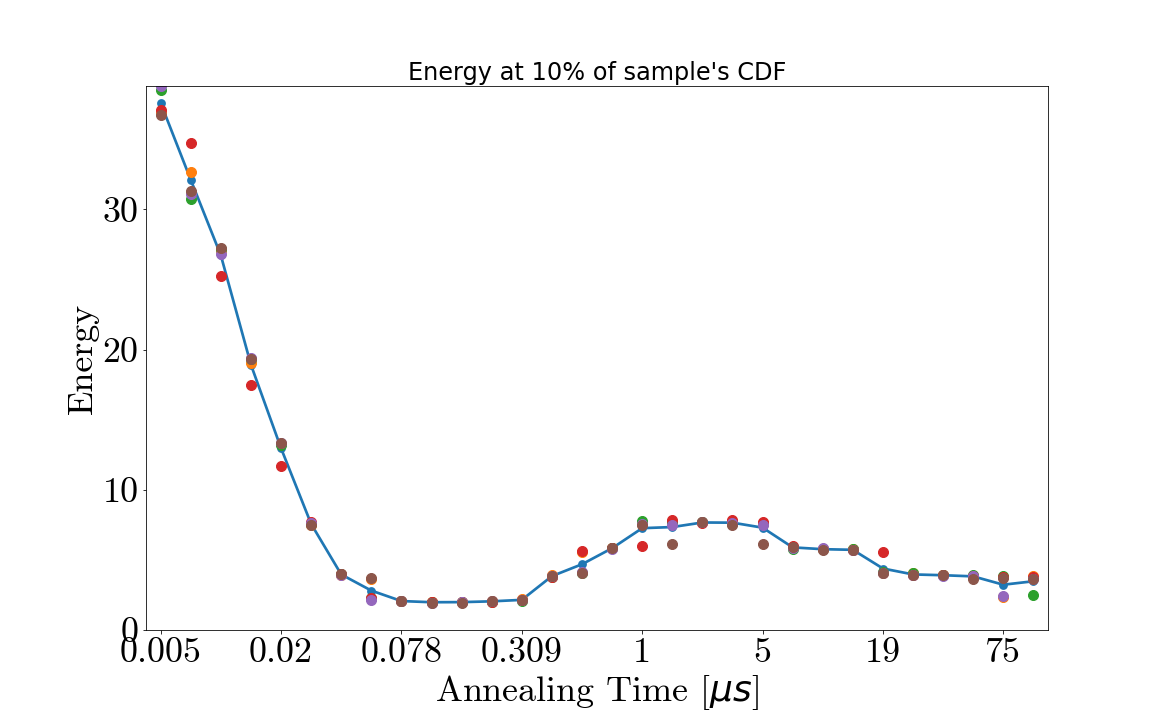}
\includegraphics[width=0.45\columnwidth]{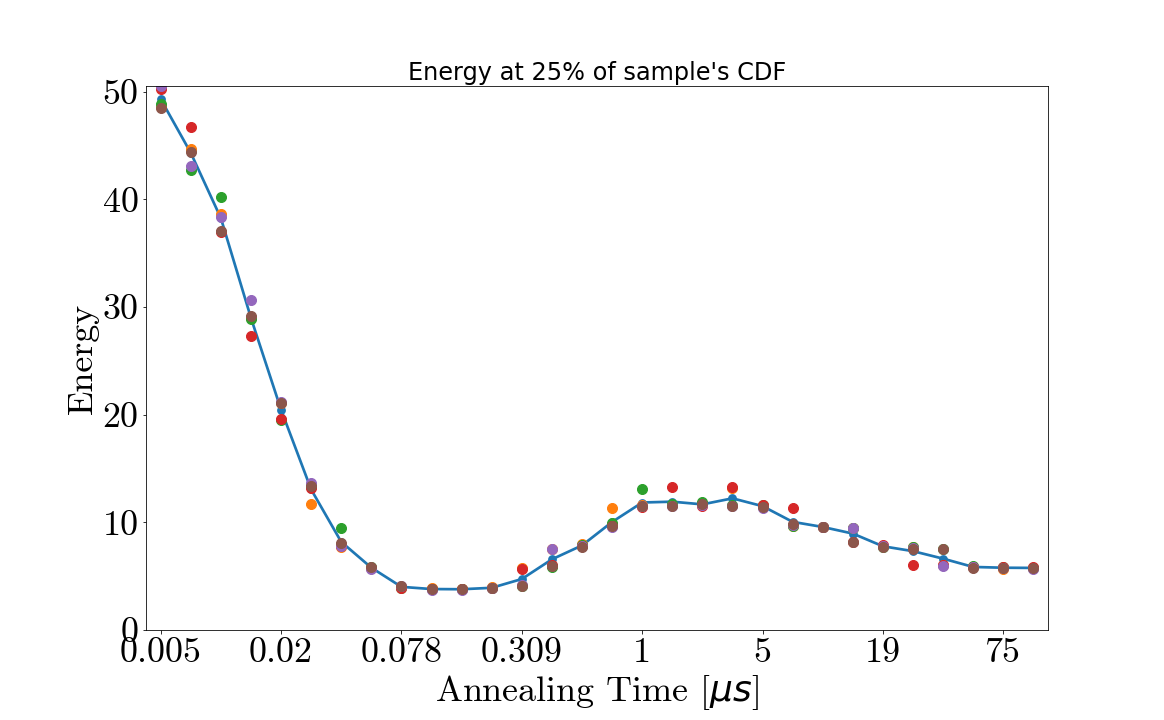}\\
Below additional plots of energy dependence on annealing time for different samples are presented:\:\\
\includegraphics[width=0.45\columnwidth]{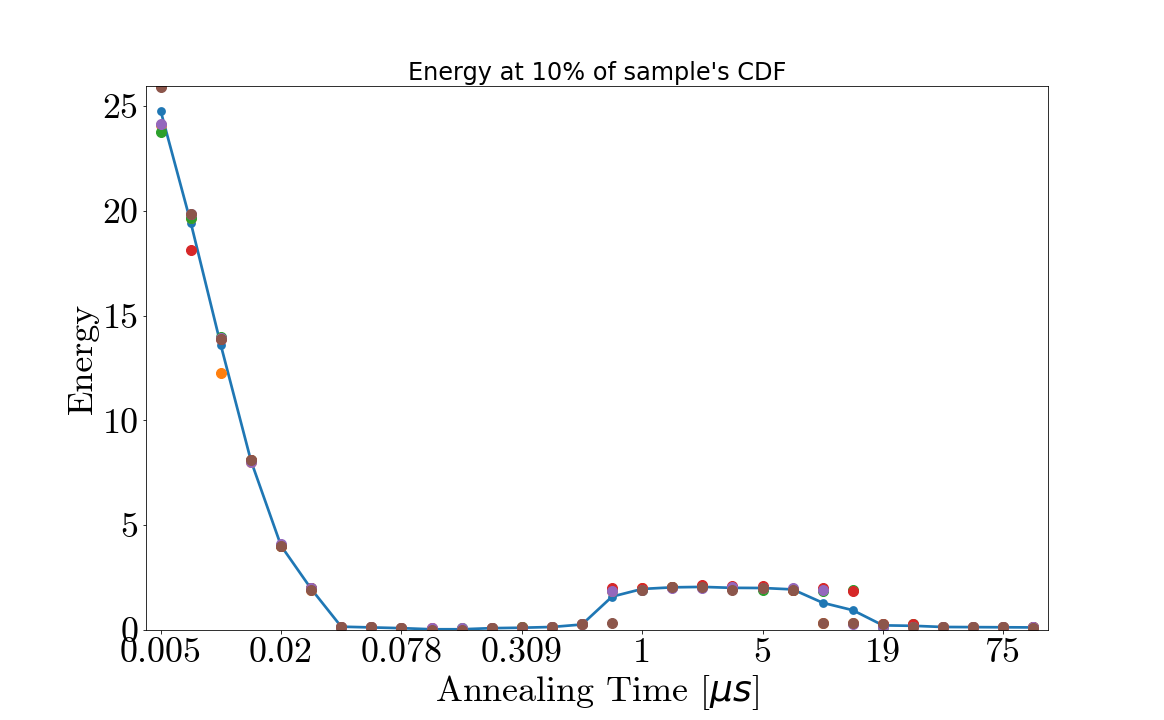}
\includegraphics[width=0.45\columnwidth]{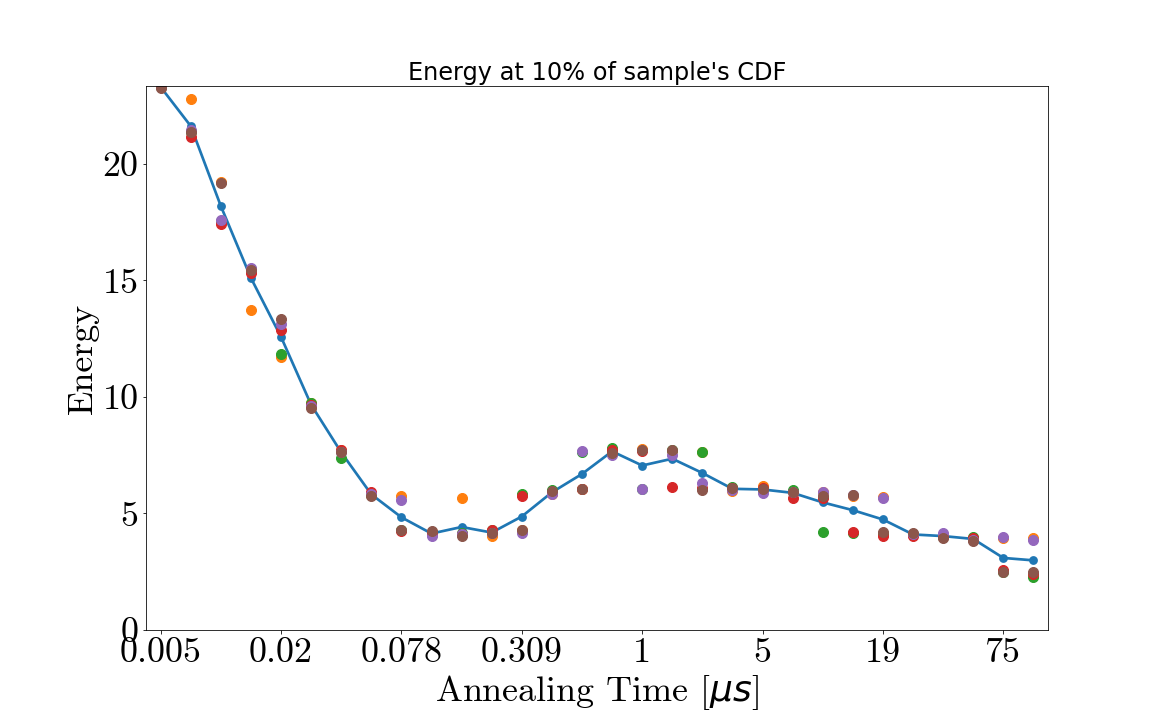}\\
\includegraphics[width=0.45\columnwidth]{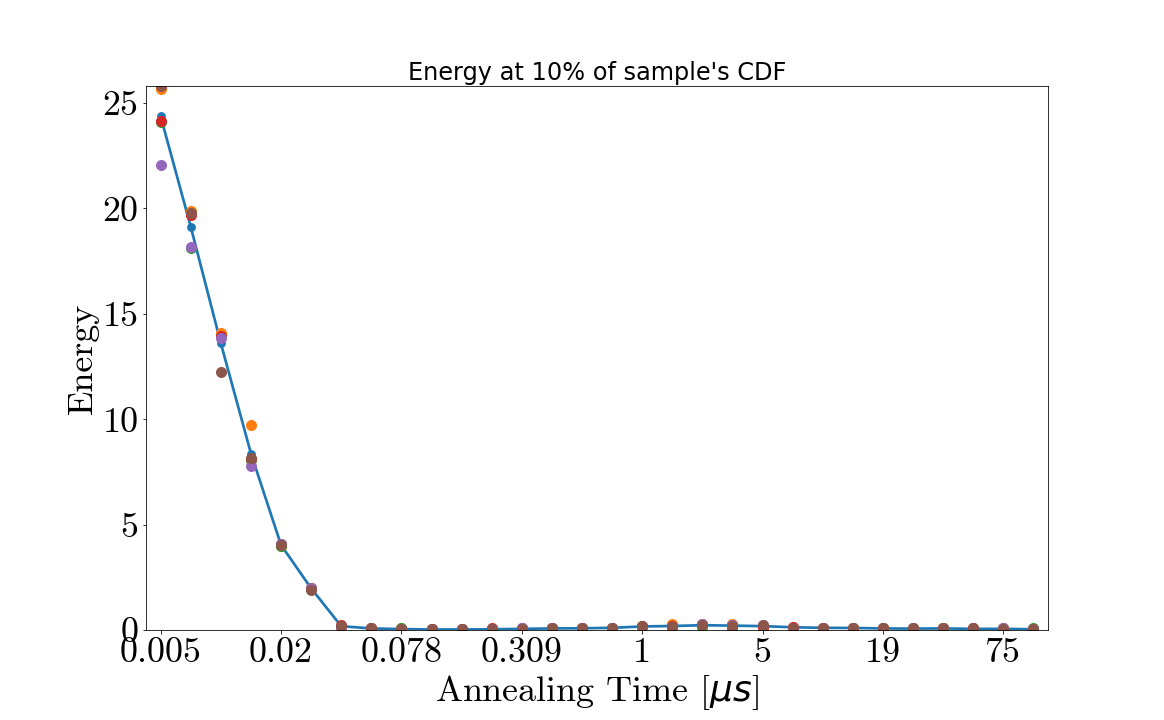}
\includegraphics[width=0.45\columnwidth]{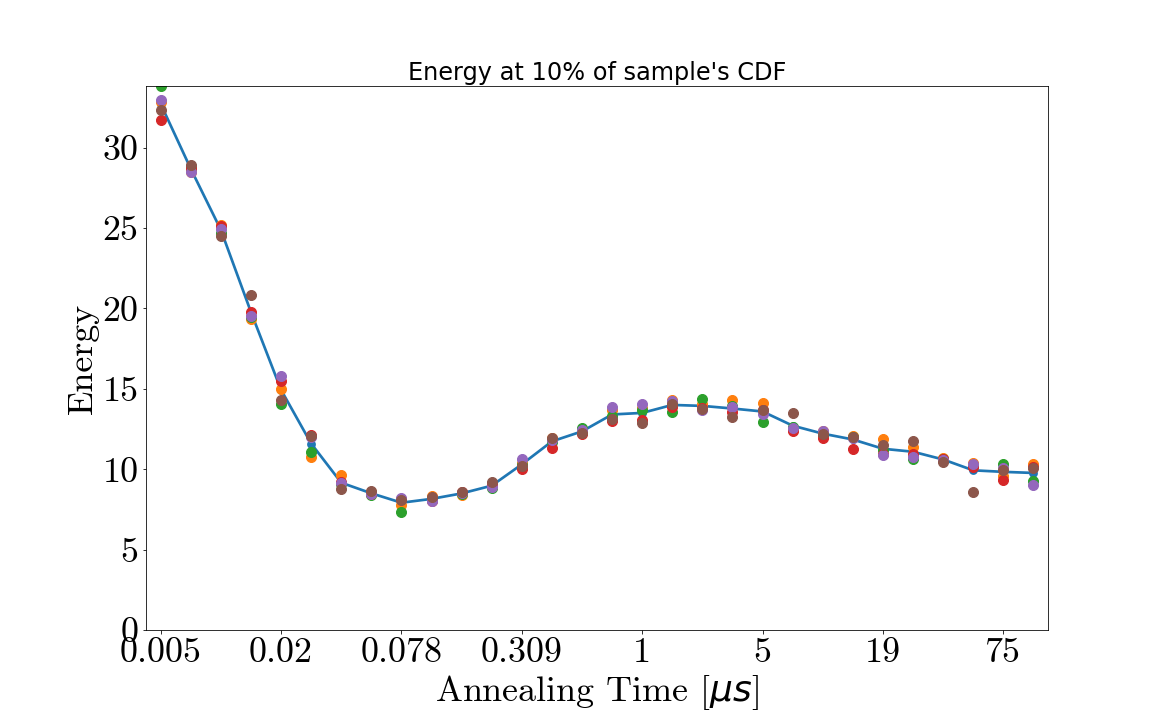}\\
\includegraphics[width=0.45\columnwidth]{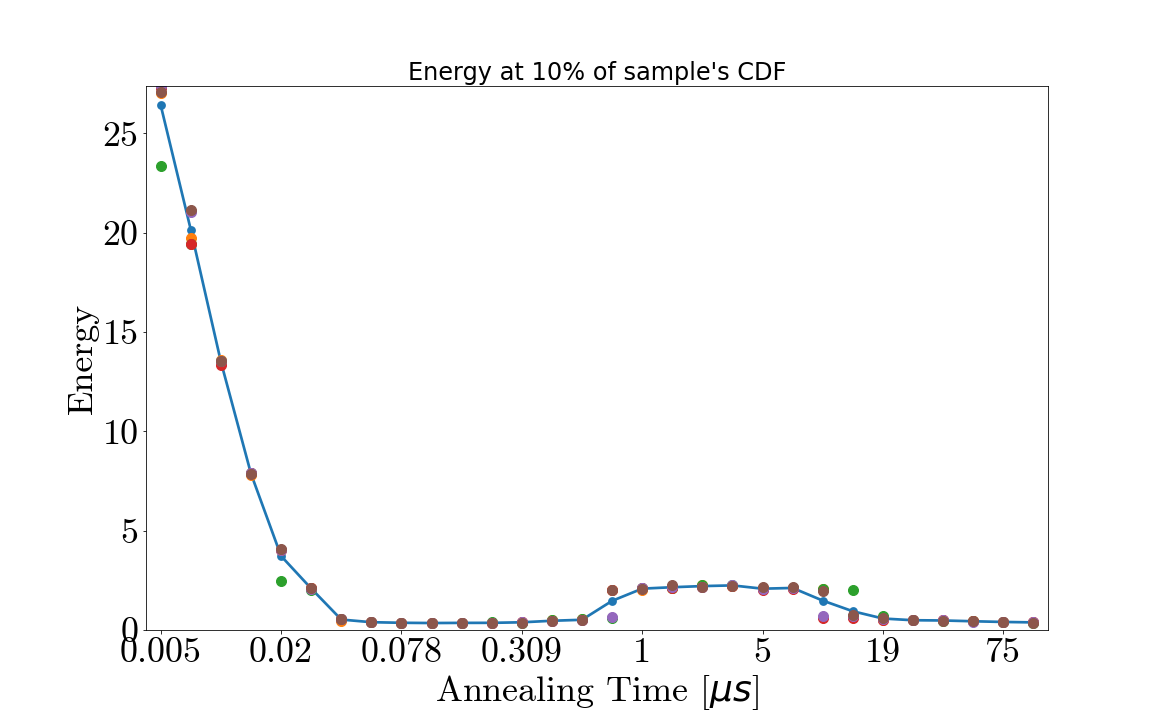}
\includegraphics[width=0.45\columnwidth]{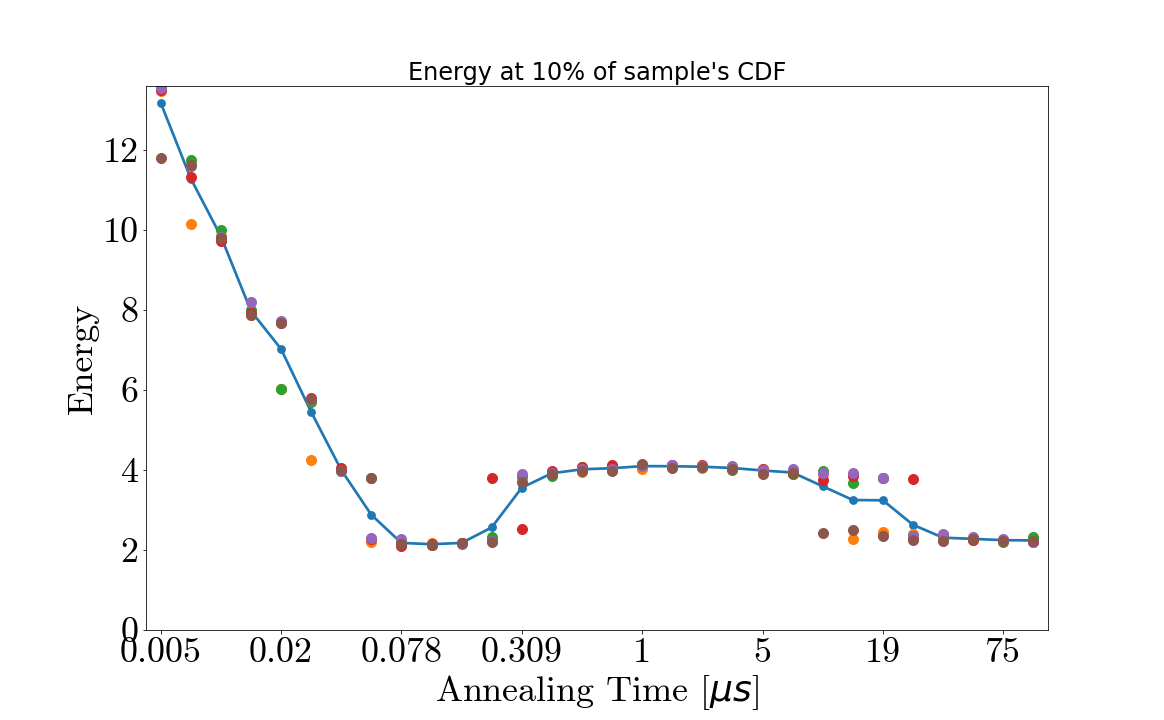}\\

\end{document}